\documentclass[lettersize,journal]{IEEEtran}
\usepackage{cite}
\usepackage{stfloats}
\usepackage{amsmath,amssymb,amsfonts}
\usepackage{graphicx}
\usepackage{textcomp}
\usepackage{xcolor}
\usepackage[hyphens]{url}
\usepackage{fancyhdr}
 \definecolor{refcolor}{HTML}{8b00ff}
\usepackage[bookmarks=true,breaklinks=true,letterpaper=true,colorlinks,citecolor=refcolor,linkcolor=refcolor,urlcolor=refcolor]{hyperref}
\definecolor{darkred}{RGB}{150,0,0}
\definecolor{darkgreen}{RGB}{0,100,0}
\definecolor{darkblue}{RGB}{80,80,255}

\newcommand{\secref}[1]{Sec.~\ref{#1}}
\newcommand{\figref}[1]{Fig.~\ref{#1}}

\newcommand{\revision}[1]{\textcolor{black}{#1}}  
\newcommand{\revise}[1]{\textcolor{black}{#1}}  
\usepackage{caption}
\usepackage{amsmath,amsfonts}
\usepackage{graphicx}
\usepackage{textcomp}
\usepackage{bbding}
\usepackage{pifont}
\usepackage{multicol}

\usepackage{booktabs}
\usepackage[binary-units=true]{siunitx}
\usepackage{array,multirow}
\usepackage{hyperref}
\usepackage{amsmath}
\usepackage{enumitem}
\usepackage{graphicx}
\usepackage{float}
\usepackage[symbol, hang,flushmargin]{footmisc}
\usepackage{bm}
\usepackage{subfigure}

\usepackage{tablefootnote}

\usepackage[noend]{algpseudocode}

\usepackage[ruled,linesnumbered]{algorithm2e}

\SetKwComment{Comment}{$\triangleright$\ }{}
\newlength\mylenin

\let\oldnl\nl % Store \nl in \oldnl
\newcommand{\nonl}{\renewcommand{\nl}{\let\nl\oldnl}} % Remove line number for one line

\newlength\mylenout

\newcommand{\tabref}[1]{Table~\ref{#1}}

\newcolumntype{L}[1]{>{\raggedright\let\newline\\\arraybackslash\hspace{0pt}}m{#1}}
\newcolumntype{C}[1]{>{\centering\let\newline\\\arraybackslash\hspace{0pt}}m{#1}}
\newcolumntype{R}[1]{>{\raggedleft\let\newline\\\arraybackslash\hspace{0pt}}m{#1}}

\usepackage{soul}
\usepackage{xurl}
\usepackage{xspace}
\usepackage{tablefootnote}

\title{SD-Acc: Accelerating Stable Diffusion through Phase-aware Sampling and Hardware Co-Optimizations} 
\author{
\IEEEauthorblockN{Zhican Wang, Guanghui He$^{*}$, \textit{Member, IEEE}, Hongxiang Fan}
\thanks{
\indent Zhican Wang and Guanghui He are with the State Key Laboratory of Micro-Nano Engineering Science, and School of Integrated Circuits, Shanghai Jiao Tong University, China (e-mail: \{wang\_zhican, guanghui.he\}@sjtu.edu.cn). Hongxiang Fan is with Imperial College London, UK (e-mail:hongxiang.fan@imperial.ac.uk).}
}
\begin{document}
\maketitle
% \pagestyle{plain}

%%%%%% -- PAPER CONTENT STARTS-- %%%%%%%%
\begin{abstract}
The emergence of diffusion models has significantly enhanced the capabilities of generative AI, leading to improved quality, realism, and creativity in image and video generation. 
Among these, Stable Diffusion (\textit{StableDiff}) become one of the most influential diffusion models for text-to-image generation, serving as a critical component in next-generation multi-modal algorithms.
Despite their advances, \textit{StableDiff} imposes substantial computational and memory demands, affecting inference speed and energy efficiency.
To address these hardware performance issues,
we identify three key challenges: \textit{1)} intensive and potentially redundant computation. \textit{2)} heterogeneous operators involving both convolutions and attention mechanisms. \textit{3)} widely varied weight and activation sizes. 

Thus, we present \textit{SD-Acc}, a novel algorithm and hardware co-optimization solution:
At the algorithmic level, we observe that high-level features exhibit high similarity in certain phases of the denoising process, indicating \revision{the potential of approximate} computation. 
Therefore, we design an adaptive \textit{phase-aware sampling} framework to reduce the computational and memory requirements.
This general framework automatically explores the trade-off between image quality and algorithmic complexity for any given  \textit{StableDiff} models and user requirements.
At the hardware level, we introduce an \textit{address-centric} dataflow to facilitate the efficient execution of heterogeneous operators within a simple systolic array. \revision{The bottleneck of nonlinear operations is comprehensively solved by our novel \textit{2-stage streaming computing} and reconfigurable} vector processing unit.
We further enhance hardware efficiency through an adaptive dataflow optimization that incorporates dynamic reuse and fusion techniques tailored for \textit{StableDiff}, leading to significant memory access reduction.
Across multiple \textit{StableDiff} models, our extensive experiments demonstrate that our algorithm optimization can achieve up to $3$ times reduction in computational demands while preserving the same level of image quality across various models.
Together with our highly optimized hardware accelerator, our approach can achieve higher speed and energy efficiency over CPU and GPU implementations.
\end{abstract}

% $14.2\sim25.6\times$ and $2.7\sim6.0\times$ energy saving over Intel 5220R CPU and NVIDIA RTX 2080 Ti GPU.

\section{Introduction}

{
Recent advancements in diffusion models~\cite{ho2020denoising} have expanded the potential of AI-generated content (AIGC), enabling a wide range of applications, such as image editing, image super-resolution, and video synthesis. Among various AIGC applications, text-to-image generation, particularly through the use of Stable Diffusion (\textit{StableDiff})~\cite{rombach2022high}, has gained considerable attention for its ability to produce high-quality images based on user-provided textual prompts. Along with this popularity, deploying models on edge devices, such as personal computers and smartphones, to meet data privacy and model customization requirements is increasingly prioritized. However, the discrepancy between the limited resources of edge devices and the significant computational and memory demands of these models hinders their wider adoption~\cite{li2024snapfusion, ma2023deepcache,wimbauer2024cache}.

To address the hardware performance challenges of these text-to-image models, previous research has primarily focused on minimizing computational demands and memory footprint at the algorithmic level, including efficient denoising sampling~\cite{lu2022dpm}, knowledge distillation~\cite{meng2023distillation}, pruning~\cite{fang2024structural}, and quantization~\cite{he2024ptqd}. Nevertheless, existing approaches suffer from several drawbacks. First, many of these algorithmic optimizations~\cite{meng2023distillation,fang2024structural,he2024ptqd} typically require time-consuming retraining or fine-tuning processes, which require significant computational resources. Second, these methods primarily focus on algorithmic optimizations without considering the dataflow characteristics and execution efficiency of the underlying hardware. Consequently, it is challenging to translate the theoretical computation reductions into practical performance gains due to potential irregularities and workload imbalances.

% where the theoretical computational reduction is hard to transfer into performance improvement due to potential irregularities and workload imbalance.
% On one hand, translating computation reduction into performance improvement
% is challenging due to potential irregularities and workload imbalance. On the other hand, deploying the model on commercial servers has less energy efficiency and potential data privacy deficiencies. 
%\tabref{tb:comp_prev_work} presents a comparison of existing acceleration solutions versus our proposed approach. 
To address the above limitations,
this paper identifies three key challenges and proposes algorithm and hardware co-optimization to jointly improve the inference speed and energy efficiency of \textit{StableDiff}.

\begin{figure*}[hbtp]
\vspace{-10pt}
\centering
\includegraphics[width=180mm]{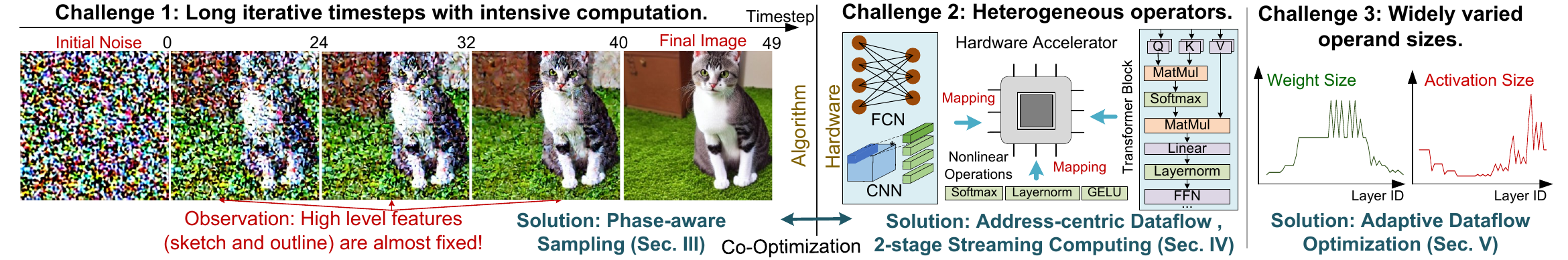}
\vspace{-5pt}
\caption{{Software and hardware challenges of the \textit{StableDiff} acceleration.}}
\label{challenge}
 \vspace{-15pt}
\end{figure*}

\textbf{At the algorithmic level}, as shown in \figref{challenge} (left), the generation process of \textit{StableDiff} starts from the random noise. The model then iteratively generates new images with progressively less noise through a time-consuming denoising process, ultimately producing the target image. 
We identify the \textbf{first challenge} as \textit{long iterative timesteps with intensive and potentially redundant computation.} 
To tackle this challenge, we first conduct a quantitative analysis of the activation distribution during the denoising process. Our analysis indicates that the denoising process can be divided into two distinct phases: \textit{sketching} and \textit{refinement}. This division into two phases is inspired by the human artistic painting process: sketching geometric contours and spatial arrangements, followed by focused refinement of local textures and details. During the \textit{sketching phase}, statistical analysis reveals that both contextual and textural information vary significantly in diffusion models, warranting the retention of the complete sampling. Conversely, the \textit{refinement phase} shows a stabilization in global and contextual information, with only local texture information undergoing active variation. The denoising process illustrated in \figref{challenge} (left) supports this observation: the contour and sketch of the cat are largely determined after approximately half of the process. While many prior works, such as \cite{ma2023deepcache, wimbauer2024cache,cambriconD2024isca}, have already identified the redundancy between consecutive timesteps, our \textit{unique motivation} is that \textit{denoising from the sketch} may require significantly fewer computations than \textit{denoising from scratch}, but the steps after a half are equally significant that cannot be early terminated. \revise{\figref{algo-subsec-profillng} provides a more specific illustration, demonstrating the variation in information levels throughout the denoising process. The darker lines indicate higher-level information (sketch), while their magnitude represents the variation. It can be observed that all levels of information change dramatically at the beginning, akin to the \textit{sketching phase}, while higher levels stabilize in the later phase, resembling the \textit{refinement phase}. A more precise definition and quantitative analysis are provided in \secref{algo-subsec-unet}.} Motivated by this observation, we reuse features from the sketching phase and preserve only a minimal number of blocks during the refinement phase. This innovative method, termed \textit{phase-aware sampling}, eliminates the need for retraining or fine-tuning while achieving significant computation reduction, maintaining comparable algorithmic performance to the original model. To \textit{generalize the proposed approach}, we propose an optimization framework that dynamically adjusts hyper-parameters according to user constraints, enabling the exploration of the trade-off between hardware performance and image quality for various \textit{StableDiff} models.

\textbf{At the hardware level,} previous co-designed accelerators are efficiently tailored for a specific algorithm but sacrifice generality, which restricts their applicability for general-purpose edge devices. To this end, our key focus is achieving efficiency while maintaining hardware simplicity and generality. Targeting \textit{StableDiff}, efficient hardware support is required for both convolution and attention operations. However, existing dedicated accelerators are optimized for either CNNs or Transformers, but they fail to efficiently support both types of operations. Specifically, as shown in \figref{challenge} (center), \textit{heterogeneous operators} pose the \textbf{second challenge}. \textit{For linear operations}, such as convolution and matrix multiplication (Matmul), current CNN accelerators \cite{chen2016eyeriss,chen2019eyeriss,nvdla} can efficiently support convolution but exhibit inefficiency in handling attention computations, and Transformer accelerators also lack the efficiency for convolution. Simply combining different engines may lead to hardware underutilization. An \textit{alternative solution} is dataflow-flexible accelerators, such as \cite{kwon2018maeri, qin2020sigma, tong2024feather}; however, this flexibility incurs significant overhead. For instance, \cite{kwon2018maeri} requires $47\%$ more area than a systolic array to efficiently support both convolution and Matmul. Another common solution is to use \textit{im2col} to transform convolution into matrix multiplication, which has seen successful deployment in both industry \cite{liao2019davinci, jouppi2017datacenter} and academia \cite{genc2021gemmini}. However, as reported in \cite{genc2021gemmini, soltaniyeh2022accelerator}, \textit{im2col} in software can account for up to $30\%$ of the end-to-end latency and result in a significant increase in memory access. Placing a dedicated \textit{im2col} hardware module can alleviate this issue. However, it still faces bank conflicts due to irregular memory access \cite{soltaniyeh2022accelerator}. Additionally, frequent conversions between formats introduce explicit latency, which is further aggravated by varying feature map shapes, kernel sizes, and strides. As a result, an elegant and low-cost solution is still lacking.

\textit{For nonlinear operations}, as widely reported \cite{kim2023stackoptimizationtransformerinference, spatten, 9586134, narechania2021vitality}, although nonlinear operations such as softmax, layernorm, and GELU constitute a small portion of the overall operations, they contribute to up to 30\% of latency on CPU/GPU platforms. This substantial latency is caused by two main inefficiencies: \textit{(i)} Softmax and layernorm require multiple passes over the data, with each pass introducing significant latency (e.g., calculation of mean, variance, and normalization in layernorm), and \textit{(ii)} they require that all related data be preloaded before computation, obstructing linear operations and presenting storage challenges. \textit{Prior works} \cite{kim2023stackoptimizationtransformerinference, kao2023flat,dao2023flashattention2fasterattentionbetter} adopt a store-then-compute strategy by fusing these nonlinear operations with preceding linear computations. Although this approach reduces off-chip access, challenges in data dependency (e.g. softmax), limited on-chip memory, and reduced parallelism can potentially degrade performance~\cite{kim2023stackoptimizationtransformerinference}. \textit{Another line} of research focuses on designing specialized and efficient engines to approximately implement nonlinear operations \cite{yu2021nnlutneuralapproximationnonlinear, 9586134, kim2021ibertintegeronlybertquantization}. However, although these circuits maintain accuracy for specific tasks, they may not perform as well in other scenarios. Moreover, dedicated engines are not flexible enough to support other nonlinear operations, and designing multiple specialized engines for various tasks can diminish overall efficiency.

% Similarly, a conventional systolic array (SA) is efficient for matrix multiplication (Matmul) of attention operations but performs poorly with CNN.
% Simply combining different engines may lead to hardware underutilization. One alternative approach is to use explicit or implicit image-to-column (\textit{im2col}) operations to transform convolution into Matmul~\cite{liao2019davinci,genc2021gemmini,zhou2021characterizing}. However, performing \textit{im2col} on the CPU can account for up to $30\%$ of end-to-end latency~\cite{liao2019davinci,genc2021gemmini}. 
% Although a dedicated \textit{im2col} hardware engine can reduce latency, it introduces high overhead due to irregular memory access. For instance, for a $3\times3$ convolution, \textit{im2col} increases the bandwidth requirement up to $9\times$ compared with Matmul, resulting in higher area and power consumption \cite{soltaniyeh2022accelerator}.
% Designing dedicated \textit{im2col} acceleration engines can alleviate this bottleneck, but they do not eliminate irregular memory access. These engines require $9\times$ the bandwidth for a $3\times3$ CNN compared to Matmul, consuming more area and power. Moreover, this approach may lack generality for convolutions with different kernel sizes or strides, which are present in \textit{StableDiff}. 

To \textit{comprehensively and holistically} solve the above issues, we employ a general-purpose systolic array (SA) as the computation core for uniform support for linear operations, achieving hardware simplicity and generality. For linear operations, we propose a reconfigurable vector processing unit to uniformly support various nonlinear operations with a single engine cost while maintaining accuracy. To enhance the hardware efficiency for \textit{linear operations}, we propose an \textit{address-centric} dataflow for convolution, which comprehensively mitigates memory irregularity and bandwidth bottlenecks due to convolution.
With our enhancements, convolutions with various kernel sizes ($1\times1$ and $3\times3$) and strides ($1$ and $2$) presents in \textit{StableDiff} are effectively supported, which enables SAs to efficiently support both convolution and attention operations with negligible extra overhead. For \textit{nonlinear operations}, to tackle inefficiency-\textit{(i)}, we divide the softmax and layernorm operations into two distinct stages. We observe that softmax and layernorm consistently occur between the preceding (pre-) and succeeding (post-) Matmul stages, and the SA writes the results of the pre-Matmul and reads the operands of the post-Matmul in a streaming manner. Therefore, we insert these distinct stages into the mandatory data streaming process, which operate in parallel with SA computations. To resolve inefficiency-\textit{(ii)}, we propose tile-decoupled scheduling via mathematical transformations, enabling softmax/layernorm computation once the first tile is generated. Our solution effectively hides the latency of nonlinear operations, rendering them transparent to the Matmul process, which is termed as \textit{2-stage streaming computing}.

By further profiling the memory traffic of \textit{StableDiff}, we observe the \textbf{third challenge}, as shown in \figref{challenge} (right): \textit{StableDiff} contains multiple downsampling and upsampling blocks, performing feature transformations across different resolutions and channel sizes, leading to \textit{widely varied} computational and memory requirements across various blocks. Building upon the address-centric dataflow, tailoring specifically for \textit{StableDiff}, we propose adaptive dataflow optimizations that apply dynamic reuse and fusion strategies across different layers to minimize off-chip access. In summary, we make the following contributions:

\begin{itemize}
    \item We observe an interesting \textit{phase division} phenomenon and propose a novel \textit{phase-aware sampling} approach to reduce the amount of computation and memory required by \textit{StableDiff}, \textit{achieving significant computational savings from an algorithmic perspective.} (\secref{sec:algo})
    
    \item We propose a \textit{address-centric} dataflow for convolution, as well as a novel \textit{2-stage streaming computation} technique for nonlinear operations, together with a reconfigurable vector processing unit. Through dataflow optimization, convolution and attention are uniformly and efficiently supported on a simple systolic array, with negligible overhead, \textit{achieving both generality and simplicity.} (\secref{sec:hw})
    
    \item We introduce the \textit{adaptive} reuse and fusion method, which dynamically applies different reuse and fusion strategies to accommodate varying weight-activation proportions across layers, thereby minimizing off-chip access, and \textit{achieving specificity and adaptability tailored for \textit{StableDiff}.} (\secref{sec:dataflow})
\end{itemize}

\section{Background}
\subsection{\revise{Preliminaries}}
Inspired by non-equilibrium thermodynamics~\cite{sohl2015deep},
diffusion models employ a sequence of timesteps to gradually generate the target images via an iterative denoising mechanism.
Starting from an initialized random noise $\mathbf{x}_{T}$ sampled from a Gaussian distribution $\mathcal{N}(\mathbf{0},\mathbf{I})$,
diffusion models iteratively produce the target image $\mathbf{x}_{0}$ through $T$ steps of denoising steps.
At each timestep $t-1$,
the output $\mathbf{x}_{t-1}$ is derived from a sampling function $\mathcal{F}(\mathbf{x}_{t}, t, \epsilon_\theta\left(\mathbf{x}_t, t, e\right))$, where $\epsilon_\theta (\mathbf{.})$ represents a neural network to predict the noise and $e$ denotes the conditional embeddings.
The specific form of $\mathcal{F}(\mathbf{.})$ varies across different diffusion models.

% sampled from a conditional probabilistic distribution modeled by previous outputs $\mathbf{x}_{t}$ and a neural network
% $\epsilon_\theta\left(\mathbf{x}_t, t\right)$ with $\mathbf{x}_{t}$ as input, which ca

Among various diffusion models,
\textit{StableDiff}~\cite{rombach2022high} emerges as a leading text-to-image diffusion model due to its excellent algorithmic performance and efficiency.
The inference of \textit{StableDiff} consists of three parts:  \textit{i)} a text encoder that generates conditional embeddings $e$ based on users' prompts, facilitating the adaptation to varied textual inputs, and \textit{ii)} a network (eg. \textit{U-Net}) to predict the noise and perform the iterative denoising process, \textit{iii)} a variational autoencoder (\textit{VAE}) to map pixels from latent space to image. 
These components work in a pipeline manner to produce high-quality images from textual descriptions.

\begin{figure}[hbt]
\centering
\includegraphics[width=88mm]{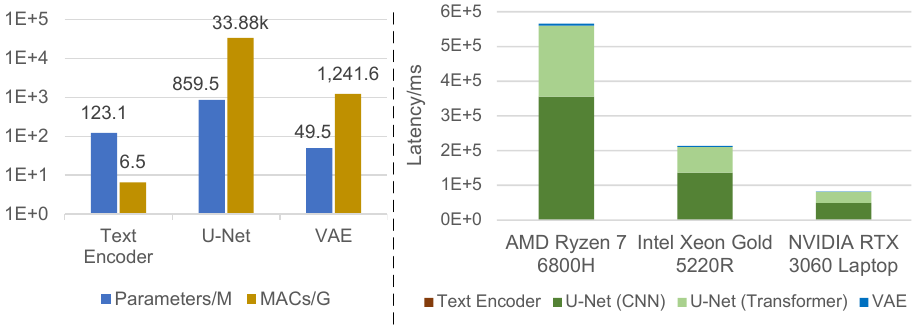}
\vspace{-10pt}
\caption{Profiling results of \textit{StableDiff} v $1.4$ with $50$ timesteps using \textit{thop} tool~\cite{thop} (left). One addition and one multiplication count as one MAC operation. \revise{The latency breakdown (right) is measured on the single-precision model.}}
\label{fig:background-profiling}
\vspace{-5pt}
\end{figure} 

To analyze the performance bottleneck of \textit{StableDiff}, we perform profiling of three main components with respect to the number of parameters, operation counts, and latency performance on both CPU and GPU platforms. As shown in \figref{fig:background-profiling} (left), among the three modules, the \textit{U-Net} parameters occupy a major portion with $860$ million parameters. Since \textit{U-Net} requires multiple executions, while \textit{VAE} and text encoder require only once, the MAC of \textit{U-Net} far outweighs those of the text encoder and \textit{VAE}. In various platforms, from \figref{fig:background-profiling} (right), it can be observed that the total inference time of \textit{StableDiff} consumes minutes of latency even on the GPU, while the CPU is slower and can take up to ten minutes. This low latency motivates our dedicated hardware and algorithm optimizations. In terms of latency breakdown, \textit{U-Net} accounts for nearly $100\times$ the latency of \textit{VAE}, while the latency of the text encoder is minimal. \revise{Within \textit{U-Net}, CNNs consume around $60\%$ of latency, and Transformers consume the rest. Due to the dominance of \textit{U-Net}, this work focuses on studying the acceleration of \textit{U-Net}, and optimize both CNNs and Transformers.}

\subsection{Network Architecture of U-Net}
\begin{figure}[htbp]

\centering
\includegraphics[width=88mm]{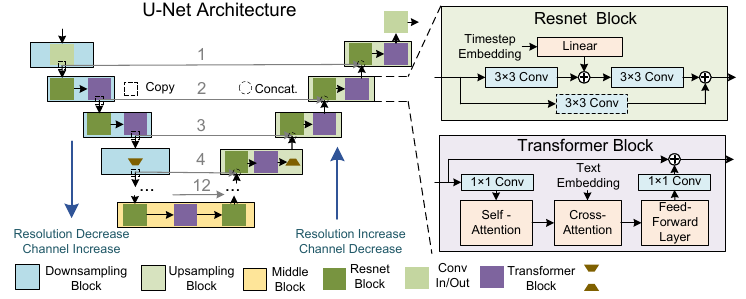}
\vspace{-5pt}
\caption{The \textit{U-Net} architecture of \textit{StableDiff} (left) with activation variations and information flow (right).}
\label{unet}
\vspace{-15pt}
\end{figure}

As shown in \figref{unet}, \textit{U-Net} consists of cascaded downsampling and upsampling blocks (both indexed from top to bottom), with a middle block between them. Common upsampling and downsampling blocks further contain cascaded ResNet and Transformer blocks. The ResNet block consists of $3\times3$ convolution layers with skip connection and incorporates time embedding information. The Transformer block incorporates $1\times1$ convolution layers and attention layers to include text-encoded information. Some blocks are slightly different, for instance, the first downsampling block only contains one $3\times3$ convolution layer, the $4_{th}, 7_{th}, 10_{th}$ downsampling block consists solely of a downsampling operation, and the $4_{th}, 7_{th}, 10_{th}$ upsampling block includes an additional upsampling operation. Downsampling and upsampling operations realize resolution variation, while the others only vary the channel sizes. The downsampling operation is achieved by $3\times3$ convolution with a stride of $2$, while the upsampling operation is achieved by nearest interpolation. The distinguishing feature of \textit{U-Net} is the skip connection from the downsampling block to the upsampling block, which concatenates with the main branch along the channel and jointly serves as the input of the upsampling block.

% \textit{U-Net} was originally proposed for image segmentation, which can capture multiple levels of feature information \cite{ronneberger2015u}. As shown in \figref{unet} (right), downsampling blocks decrease the resolution of the original image while increasing the channel size, whereas upsampling blocks restore the image to a higher resolution with a reduced channel size. In the downsampling blocks, activations with high resolution and small channels in the shallow layers carry more pixel-level and texture information, whereas activations with low resolution and large channels in the deeper layers carry more spatial and contextual information. In the upsampling blocks, the process begins with low-resolution activations, which contain more spatial information. After being concatenated with the activation from the skip branch, it gains more pixel-level information as it progresses deeper and finally outputs comprehensive information. The multiple levels of information flow in \textit{U-Net} motivate our dedicated strategies for the denoising process of \textit{StableDiff} \revision{for approximate} computation.

\section{Algorithm Optimization}\label{sec:algo}
Our phase-aware sampling aims to \revision{explore approximate} computations in a hardware-friendly manner, leveraging the similarities and distinct characteristics during different phases of diffusion models. Drawing inspiration from the human painting process, where the early phase involves outlining the skeleton followed by a focus on details in the later phase, we observe a similar principle in the denoising process, revealed by statistic analysis of \textit{U-Net} (\secref{algo-subsec-unet}). \revise{Subsequently, based on these observations, we define the phase division phenomenon and present a detailed sampling scheme in~\secref{algo-subsec-sampling}}, which efficiently reduces computational costs without the need for retraining or fine-tuning while preserving the image quality. Lastly, we provide an optimization framework in~\secref{algo-subsec-framework} to enable the customized exploration of the trade-off between speedup and image quality.

\subsection{\revise{Key Observations}}
\label{algo-subsec-unet}

The typical human painting process begins with the sketching phase, which involves drawing geometric contours and establishing spatial arrangements. This phase is followed by a refinement stage, where local textures and details are gradually inserted.
These two phases demand different levels of attention and focus on distinct elements, motivating to explore: \textit{1)} whether the denoising process follows a similar two-phase approach, and \textit{2)} whether varying degrees of \revision{approximation} exist within these distinct phases.  
To investigate these questions, we perform a quantitative analysis of the internal activations and predicted noise levels of \textit{U-Net}, shown in~\figref{algo-subsec-profillng}.

\begin{figure}[hbtp]
\centering
\includegraphics[width=88mm]{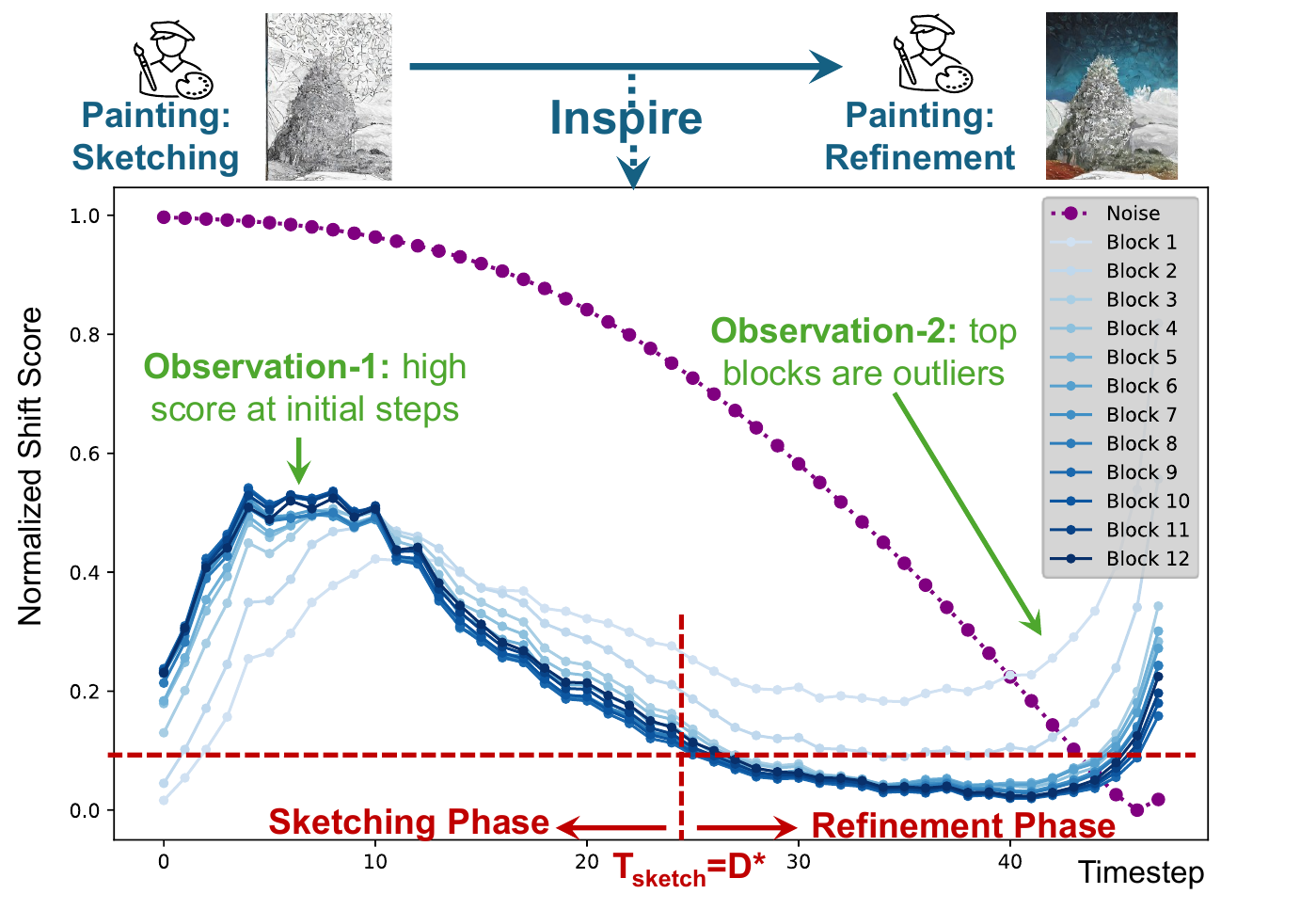}
\vspace{-10pt}
\caption{Normalized shift score of activations across the 50-timestep denoising process of \textit{StableDiff} v$1.4$, with noise and outlier curve denoted ($100\times100$ image generations). \revise{$T_{sketch}$ is set equal to $D*$ in the figure.} }
\label{algo-subsec-profillng}
\vspace{-15pt}
\end{figure} 

We define the shift score to indicate the difference in the activations between two adjacent timesteps. 
The calculation of the shift score $S$ can be formulated as
\revision{
\begin{equation}
  \textstyle
S^i_t = \frac{\Vert A^i_t-A^i_{t-1}\Vert_2}{\Vert A^i_{t-1}\Vert_2}  (t>0)
\label{shift_score}
\end{equation} }where $A^i_t$ represents the input activation from the main branch of the $i_{th}$ upsampling block at timestep $t$ and $\Vert . \Vert_2$ is the L2 norm function.
To measure the variations of the shift score across denoising timesteps, it is necessary to choose a scheduler and specify the total number of timesteps. 
For this analysis,  we randomly sample $100$ prompts from the PartiPrompts dataset~\cite{yu2022scaling} and generate $100$ images for each prompt using the 50-timestep scheduler~\cite{karras2022elucidating}. 
Given that shift scores can differ in scale across various blocks, 
we employ min-max scaling to normalize each block's scores to a uniform range of $[0,1]$, where the same scaling is also applied to the noise across different denoising timesteps.
Subsequently, the shift score is calculated for each block per timestep and averaged across different images. According to the profiling results presented in~\figref{algo-subsec-profillng}, we have the following two key observations, which have been extensively validated with various models and configurations. \revise{These observations motivate the finding of phase division and drive the exploration of phase-aware sampling (\secref{algo-subsec-sampling})}.\\
\textbf{\textit{Key Observation 1:}} \textit{U-Net} experiences varying difference across distinct denoising timesteps.
As shown in~\figref{algo-subsec-profillng},
most blocks exhibit the following pattern: the shift scores change dramatically at the beginning, rising and then falling in a wave-like manner, before stabilizing and experiencing a slight rise again near the end, indicating their proximity to the final target is nearly accomplished in the initial phase.\\
\textbf{\textit{Key Observation 2:}} Specific blocks, such as \textit{block-1} and \textit{block-2}, still exhibit high variance in the later denoising timesteps.
This behavior indicates that the upper blocks of \textit{U-Net} become important in later timesteps, supporting the hypothesis that \textit{StableDiff} focus more on detailed texture information in the final phase, similar to humans. 
In contrast, spatial and contextual information initially undergoes substantial changes but then remains stable.\\

\subsection{Phase-aware Sampling}
\label{algo-subsec-sampling}
\revise{\textbf{\textit{Phase Division:}}
Based on the above observations, this work proposes to divide the denoising process into two distinct phases:} sketching and refinement. Inspired by K-means clustering~\cite{hartigan1979k} ($K=2$), the transition timestep $D$ between the two phases can be determined by sweeping the timesteps:
\begin{equation}
  \textstyle
D^*=\underset{D=1\cdots T-2} {argmin}\ \sum_{t=0}^{D}(\bar{S_t}-\bar{{\mu}}_{skt})^2+\sum_{t=D+1}^{T-1}(\bar{S_t}-\bar{{\mu}}_{ref})^2,
\label{loss}
\end{equation}
$\bar{S_{t}}$ represents the averaged shift score of the blocks except for the outlier curves. $\bar{\mu}_{skt},\bar{\mu}_{ref}$ stands for the mean $\bar{S_{t}}$ of the divided phases. Due to distinctive characteristics, the above equation finds the global optimal transition timestep to minimize the variance sum of the two phases. \revision{$D*$ is a meta-parameter obtained through the calibration phase, which will be introduced in \secref{algo-subsec-framework}}.

\begin{figure}[t]
\centering
\includegraphics[width=88mm]{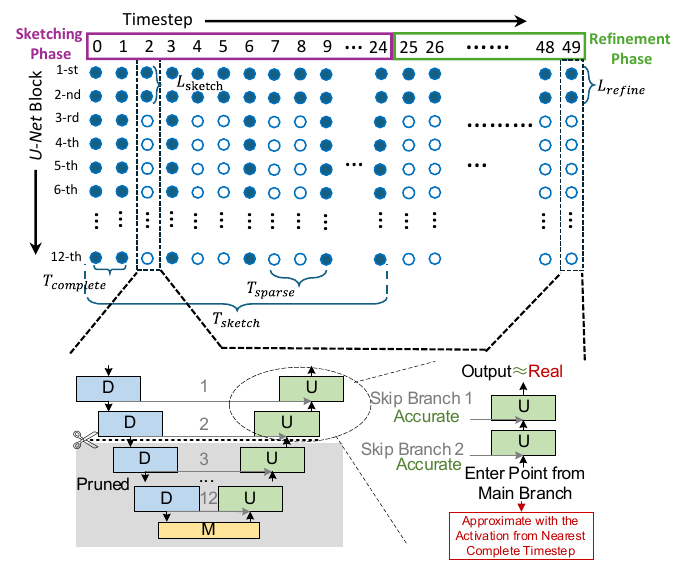}
\vspace{-10pt}
\caption{An example of phase-aware sampling with layer skipping. \revise{The pruning process of two phases is shown at the bottom.}}
\label{pruning}
\vspace{-15pt}
\end{figure} 
To explore \revision{approximate computation}, we adopt different strategies for the \revise{two phases} outlined in~\secref{algo-subsec-unet}.
\revise{As shown in \figref{pruning} (top), we define the parameter set $\{T_{sketch}, T_{complete}, T_{sparse}, L_{sketch}, L_{refine}\}$, which can be categorized into spatial and temporal types. For the spatial parameters, $L_{sketch}$ and $L_{refine}$ represent the number of top blocks in the incomplete \textit{U-Net} during the sketching and refinement phases, respectively. $T_{sketch}$, $T_{complete}$, and $T_{sparse}$ are temporal parameters, where $T_{sketch}$ refers to the duration of the sketching phase, $T_{complete}$ represents the number of timesteps reserved for the complete \textit{U-Net} at the beginning, and $T_{sparse}$ indicates the sampling period of complete \textit{U-Net} during the sketching phase.} A smaller $T_{sketch}$ indicates more aggressive compression, but it must not be less than $D^*$ to ensure stability.

Specifically, \textit{during the sketching phase}, when the shift scores are more significant and the predicted noise is large, we maintain complete runs of the entire \textit{U-Net} for the first $T_{complete}$ timesteps. Then, for the remaining $T_{sketch} - T_{complete}$ timesteps of the sketching phase, we employ uniform sparse sampling, running the complete network every $T_{sparse}$ timesteps and executing only the first $L_{sketch}$ blocks for the other timesteps to improve speed. \textit{During the refinement phase}, since only the topmost blocks exhibit high shift scores, as shown in~\figref{algo-subsec-profillng}, we retain these downsampling and upsampling blocks while pruning the execution of others. \revise{The detailed pruning process for both phases is illustrated in~\figref{pruning} (bottom). As shown in the right zoom-in subfigure, since the activation from the latest complete timestep is reused as the entry point for the retained blocks, which is close to the real value, the output approximately matches the real value.} The number of retained blocks, $L_{refine}$, is determined by the significance level of the shift score based on the profiling results and should be no less than the number of outlier blocks. Additionally, $L_{sketch}$ should be no smaller than $L_{refine}$.

\iffalse
It is worth noting that the closest previous work to ours is \textit{DeepCache}~\cite{ma2023deepcache}.
It can be seen as a subset of our method and performs a consistent sampling strategy across all timesteps without considering phase awareness. Also, the hyper-parameters related to layer skipping such as $L_{refine}$ and $L_{sketch}$ are determined adaptively using our proposed framework (\secref{algo-subsec-framework}), instead of being fixed in \textit{DeepCache}~\cite{ma2023deepcache}.
\fi

\revision{It is worth noting that layer skipping for \textit{StableDiff} is not our proprietary contribution. Deepcache \cite{ma2023deepcache} also adopts block skipping, while Cacheme\cite{wimbauer2024cache} specifically skips attention. However, they don't consider phase awareness, and the parameters, such as $L_{refine}$ and $L_{sketch}$ are fixed, rather than determined adaptively using our framework (\secref{algo-subsec-framework}).}

% According to Fig. \ref{curve}, the diffusion process can be divided into two phases: the variational phase and the frozen phase. A predefined threshold, such as $0.1$, can determine the boundary between the two phases. In the variational phase, every block significantly contributes to the entire process, while in the frozen phase, the majority of blocks remain stable, and only the shallowest blocks influence the output. This phase division suggests redundant computing if the complete U-Net is consistently executed throughout the entire process, offering an opportunity to reuse the frozen spatial and contextual information. As depicted in Fig. \ref{pruning}, during the frozen phase, only the shallowest downsampling and upsampling blocks are retained, while the other blocks can be safely pruned. For the retained blocks, the skip connections from downsampling blocks are accurately computed from the input, and the only uncertainty lies in the entry point of the main branch before the retained upsampling block. Naturally, we can reuse the frozen features at the last timestep of the variational phase, acting as the resource for approximation in this scheme, which closely resembles reality. Consequently, for the frozen phases, the majority of the U-Net is pruned, leading to a significant operations reduction in a hardware-friendly manner. 

\subsection{General Optimization Framework}\label{algo-subsec-framework}
% \begin{figure}[hbtp]
% \centering
% \includegraphics[width=88mm]{curve2.pdf}
% \caption{Shift score of activations across the 25-step diffusion process with noise and complete steps denoted.}
% \label{curve2}
% \end{figure} 

As discussed in~\secref{algo-subsec-sampling},
our proposed phase-aware sampling is defined by several hyper-parameters including $\{T_{sketch}, T_{complete}, T_{sparse}, L_{sketch}, L_{refine}\}$.
These parameters decide the trade-off between the speedup and image quality.
To generalize our approaches to different \textit{StableDiff} models and facilitate the exploration of performance trade-offs,
we propose a framework to optimize these hyper-parameters according to users' requirements.

% Our method is also compatible with acceleration by timestep reduction. As depicted in Fig. \ref{curve2}, An emergent phase division is observed even with $25$ total timesteps. In addition to adjusting the total number of timesteps, another opportunity for further reducing operations arises from the variational phase. Inspired by the frozen phase, where one activation from the last timestep of the variational phase is reused for every subsequent timestep. Consequently, we can uniformly divide the variational phase into several groups, each spanning a few timesteps. The complete U-Net is executed only at the first timestep for each group, and the activation is reused for the remaining timesteps with incomplete U-Net. In addition, an increase appears at the end of the curve, but our experiments show that keeping the entire accurate U-Net is not necessary, because the absolute value of noise is extremely small at the end. Similarly, we suggest keeping some timesteps at the beginning due to the large noise proportion. 

\begin{figure}[ht]
\centering
\vspace{-10pt}
\includegraphics[width=85mm]{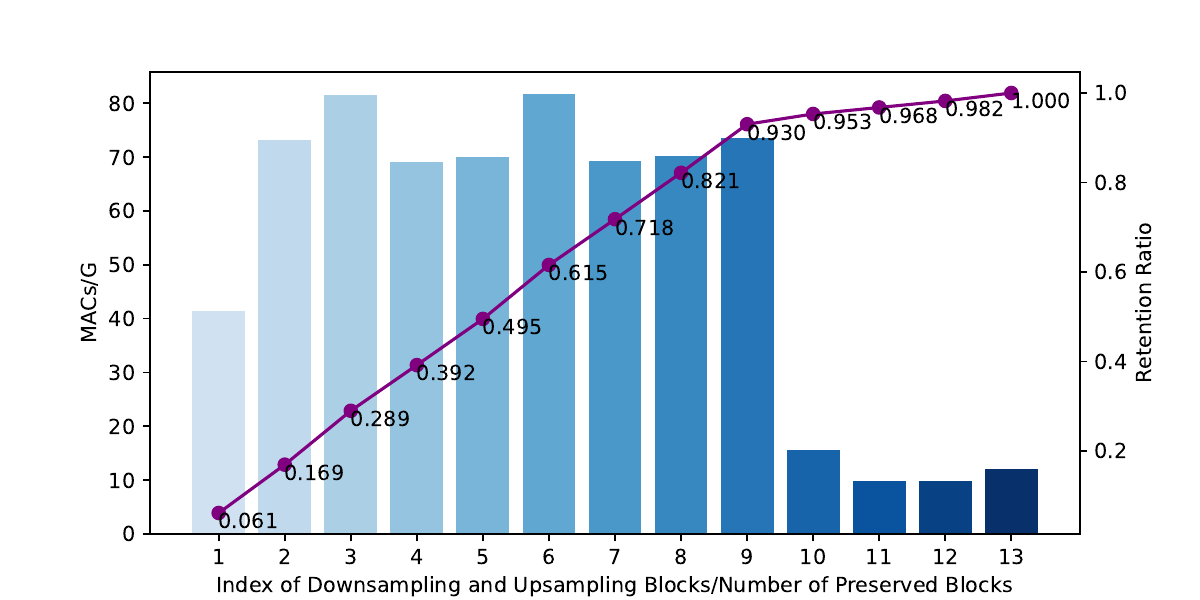}
\caption{The MAC breakdown of the downsampling and upsampling blocks in \textit{U-Net} with the cost function curve. $l=13$ denotes the entire \textit{U-Net} including the middle block.}
\label{unet-mac-per-step}
\vspace{-5pt}
\end{figure} 
\begin{figure}[ht]
\vspace{-10pt}
\centering
\includegraphics[width=88mm]
{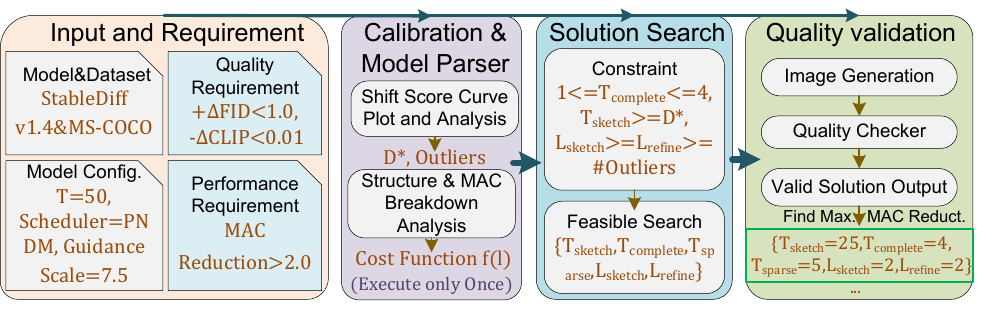}
\vspace{-10pt}
\caption{General framework for Phase-aware sampling (Examples in orange).}
\label{framework}
\vspace{-10pt}
\end{figure}

To evaluate the MAC reduction,
we build a cost function $f(l)$ that quantifies the computational overhead associated with running the first $l$ downsampling and upsampling blocks of \textit{U-Net}.
As presented in~\figref{unet-mac-per-step},
we examine the MAC operation counts 
for each block within \textit{U-Net}.
The cost function, $f(l)$, is assigned as the cumulative computational ratio for the initial $l$ blocks and normalized by the total MAC computation of the entire network, which is depicted as the purple line in~\figref{unet-mac-per-step}. 
% We use $f(-1)=1$ to indicate the complete execution of the entire \textit{U-Net}.
Thus, the overall MAC reduction across all denoising timesteps can be formulated as:
\begin{equation}
  \textstyle
MAC_{reduce} = \frac{T}{ \sum_{t=0}^{T-1}f(l_{t})},
\label{mac_reduction}
\end{equation}
where $l_{t}$ represents the number of running blocks at timestep $t$.
The primary objective of our framework is to maximize the MAC reduction in Equation~(\ref{mac_reduction}) while ensuring the image quality meets user requirements.

\figref{framework} presents an overview of our framework.
Inspired by the calibration phase in neural network post-training quantization~\cite{nagel2021white}, we also employ a calibration prompt dataset to optimize the hyperparameters offline. \revise{The calibration dataset is generated by randomly selecting $5\%$ of the targeted prompt dataset. The calibration process takes less than one hour on a V100 GPU,  and we also find that $D^*$ is quite robust to the randomness of the prompt.} Firstly, users specify the targeted model and datasets, model configurations, and the requirements of performance and image quality (\secref{subsec-exp-setup}). Secondly, the framework plots and analyzes the shift core curve to obtain the outliers and $D*$, as described in \secref{algo-subsec-unet}. The model parser analyses the model structure and the MAC breakdown to obtain the cost function. Thirdly, the framework conducts the solution search under the constraints, obtaining the feasible ones. Finally, the framework generates the images according to the solution and validates their quality. It outputs valid solutions, providing the solution with maximum MAC reduction.
Following this four-step process,
users can explore the trade-off between hardware performance and image quality tailored to their specific scenarios.

\section{Unified Hardware Architecture}\label{sec:hw}

% \textit{StableDiff} involves various distinct operators, including matrix multiplication in the attention layers, as well as convolution with varying hyperparameters, such as kernel sizes ($1\times1$ and $3\times3$) and different strides. While heterogeneous engines dedicated to convolution and attention can be leveraged, they often sacrifice generality and processing engine (PE) utilization. 
To uniformly support both convolution and attention, we propose \revise{an} \textit{address-centric} dataflow in (\secref{sec:addr}). Subsequently, as described in~\secref{Unified Mapping to the Systolic Array}, we introduce our simple and general-purpose hardware architecture with the mapping solution. In \secref{sec:nonlinear}, we introduce a novel \textit{2-stage streaming computing} strategy for nonlinear operators, and \secref{sec:VPU} illustrate the corresponding reconfigurable hardware support.%Then, we briefly introduce efficient support of high-stride convolution and deconvolution to realize downsampling and upsampling operations. 

\begin{figure*}[htbp]
% \vspace{-5pt}
\centering
\includegraphics[width=180mm]{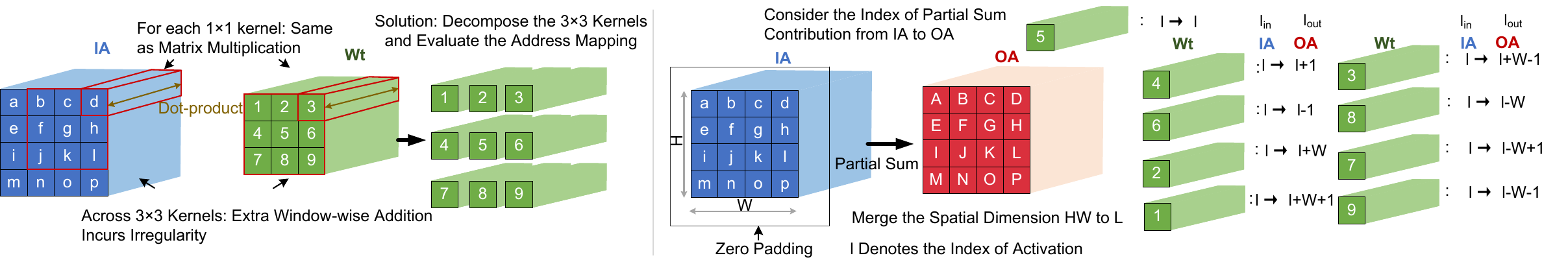}
 \vspace{-10pt}
\caption{Convolution decomposition (left). Address mapping scheme of the partial sum from input activation to output (right). }
\label{mappinglaw}
 \vspace{-10pt}
\end{figure*} 

\subsection{Address-centric Reinterpretation}
\label{sec:addr}

\begin{figure}[hbtp]
\vspace{-10pt}
\centering
\includegraphics[width=88mm]{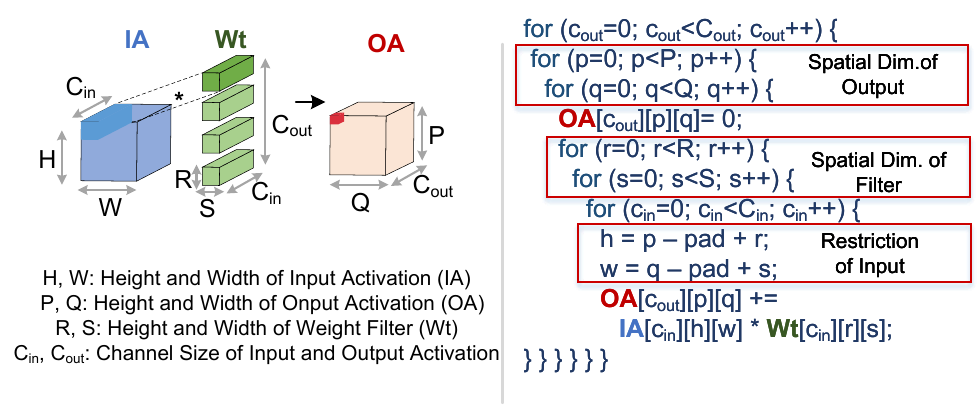}
\vspace{-10pt}
\caption{Illustration of convolution (left), conventional loop nest representation (right).}
\label{cnn}
\vspace{-10pt}
\end{figure}

\textbf{Notations and inefficiencies.} Conventionally, convolution can be represented using a six-layer loop nest in~\figref{cnn}~(right) assuming a batch size of one, with the notation $P, Q, H, W, R, S, C_{in}, C_{out}$ depicted in~\figref{cnn}~(left). We use capital letters to represent the loop border, and the lowercase to represent the variable in the corresponding dimension. However, as shown in \figref{cnn} (right), the spatial dimensions and shift window manner lead to memory access irregularity, making it challenging to efficiently support convolution on a systolic array at negligible cost.
% The conventional approach relies on \textit{im2col} operations to transform it into matrix multiplication~\cite{jia2014caffe,chetlur2014cudnn,zhou2021characterizing}, as illustrated in~\figref{cnn}~(right). Nevertheless, many accelerators are designed for matrix multiplication, making it challenging to efficiently support CNNs if \textit{im2col} is performed on software, due to the large latency consumed by irregular and discontinuous memory access and element repetition. 
% Consequently, several accelerators incorporate dedicated \textit{im2col} modules to reduce latency~\cite{liao2019davinci,genc2021gemmini}. 
% However, integrating an \textit{im2col} module, as compared to the original matrix multiplication, requires higher bandwidth. 
% The \textit{im2col} module also involves irregular and large memory accesses to prepare the matrix, leading to elevated costs and energy consumption. Therefore, we strive to offer a cost-effective solution that balances the competing demands of hardware cost and performance considerations.

\textbf{Overview.} The aforementioned issues are addressed through address-centric re-interpretation. \textit{First}, rediscover the source of irregularity in convolution compared to Matmul. A $3\times3$ convolution example is shown in~\figref{mappinglaw}~(left), it is evident that the multiplication and addition across the input channel are the same as the dot-product in the Matmul, whereas the additional addition within the window is unique to convolution. It is straightforward to decompose the convolution kernel into $1\times1$ kernels, and each $1\times1$ convolution operation becomes Matmul. \textit{Second}, re-evaluate the address relationship between input and output. These $1\times1$ convolutions only generate a partial sum compared with the original form, we identify a simple address mapping scheme between the index of input (source of the partial sum) and the output activation (destination of the partial sum). 

\textbf{Re-interpretation.} Take the most commonly used convolution, the same convolution as the example, it involves zero padding to ensure the same height and width between the input and output. The original $h, w$ dimensions are combined into one dimension $l$ and $L=HW$, $l_{in}$ $l_{out}$ denote the input and output activation. As shown in~\figref{mappinglaw}~(right), according to the different kernel positions in the original $3\times3$ convolution, address mapping differs correspondingly. For example, the center $1\times1$ kernel (denoted as \textcolor{darkgreen}{5}) convolves with input activation \textcolor{darkblue}{a} and generates the partial result, where this partial result should be added to output activation at the position \textcolor{darkred}{A}. Similarly, the partial result from \textcolor{darkblue}{b, c, d...} should be added to \textcolor{darkred}{B, C, D...}. Due to the same index between source and destination, the address mapping can be summarized as $l\xrightarrow{}l$. For kernel-\textcolor{darkgreen}{4} in~\figref{mappinglaw}, it convolves with \textcolor{darkblue}{a, b, c, d...}, and the partial result should be added to \textcolor{darkred}{B, C, D, E...}, and the address mapping is $l\xrightarrow{}l+1$. The rest can be deduced similarly as shown in~\figref{mappinglaw}~(right). 
Notably, certain edges of the input don't have corresponding output, necessitating detection and avoidance of addition for these edges, where we use a flag to denote these. 
% \textit{Benefits.} Throughout the process: \textit{1)} Memory access remains regular, incrementing monotonically for both input and output. The address mapping scheme can be easily implemented by setting the address bias of the output. \textit{2)} Bandwidth requirements are equally distributed between input and output \textit{3)} Importantly, this address mapping scheme absorbs cumbersome spatial dimensions. Parallel fetching of an input channel chunk indexed by $l_{in}$ and parallel generation of output channels chunk with $l_{out}$ are facilitated. 
% For instance, when considering kernel-\textcolor{darkgreen}{9}, its convolution with the input's leftmost column and uppermost row should not affect the output. Typically, the edge portion exhibits a minimal ratio due to the large image sizes.

\begin{figure*}[ht]
\vspace{-10pt}
\centering
\includegraphics[width=180mm]{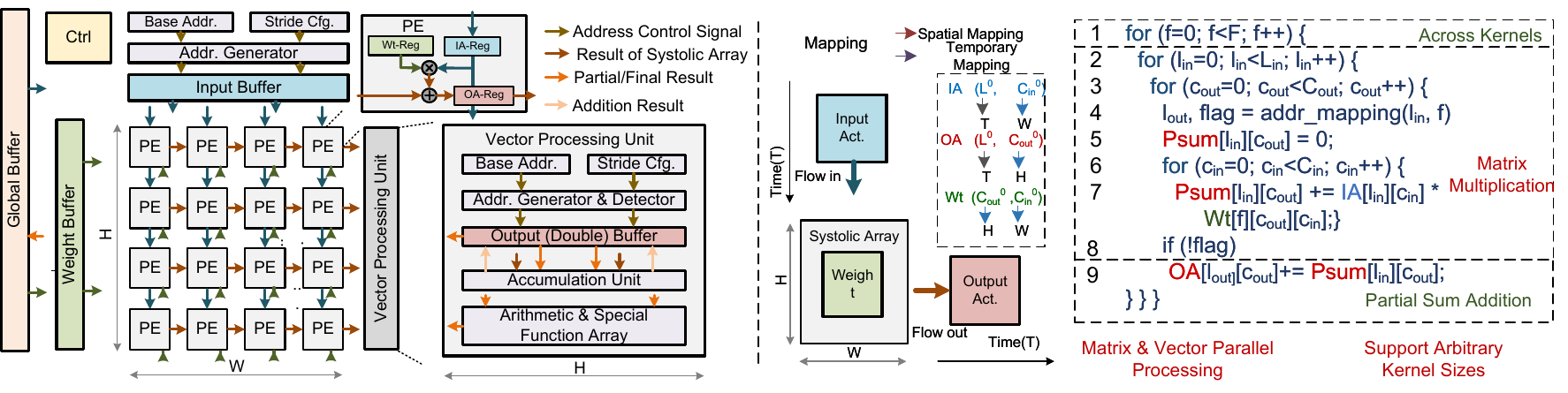}

\caption{Hardware architecture (left), mapping (center), and address-centric dataflow (right).}
\label{hardware}

\vspace{-10pt}
\end{figure*} 
\subsection{Mapping and Hardware Architecture}
\label{Unified Mapping to the Systolic Array}
\textbf{Storage Format and Address-centric Dataflow.} We define a new storage format for the address-centric dataflow. Because the address mapping scheme absorbs cumbersome spatial information, the original spatial dimensions of input $H, W$, and output $P, Q$ can be eliminated, and they can be uniformly merged into a single dimension $L$. 
Consequently, all activations can be represented as $(L, C_{in})$. 
Similarly, the spatial dimensions of weights, $R$ and $S$, can be combined as $F=RS$, resulting in weight storage as $(F, C_{out}, C_{in})$. 
As a result, our proposed dataflow can be represented using a simplified four-layer loop nest, which incorporates matrix multiplication and partial sum addition. The transformed dataflow is summarized in~\figref{hardware} (right), the entire process is referred to as an operator, \textit{Uni-conv}. Lines 2-8 correspond to matrix multiplication with the size $(L_{in}, C_{in})\times(C_{in}, C_{out}) \xrightarrow{}(L_{in}, C_{out})$. The outermost loop in Line 1 denotes the partial sum addition across each $1\times1$ kernel. 
Crucially, matrix multiplication can be performed in parallel with the partial sum addition of Line 9, effectively hiding latency.
% \begin{figure}[ht]
% \vspace{-10pt}
% \centering
% \includegraphics[width=80mm]{./figs/uni-conv.pdf}
% \vspace{-10pt}
% \caption{Dataflow of the \textit{unified convolution}.}
% \label{uni-conv}
% \vspace{-5pt}
% \end{figure} 

\textbf{Hardware Overview.} As depicted in~\figref{hardware}~(left), our architecture consists of a weight-stationary systolic array for Matmul, on-chip buffers, and a vector processing unit (VPU) for partial sum addition and nonlinear operations, which will be introduced in the next sections. The address generator produces the incremental address based on the base address and the stride configured to realize the address mapping scheme. The systolic array has a height and width of $H, W$, (irrelevant to the size of input activation), while the VPU also has a height of $H$, i.e., $H-$ parallel. Transformers are supported by mapping matrix multiplication on the systolic array, and nonlinear operations on the VPU concurrently. 
%, and we focus on the convolution in this work
Given the conflict between the limited size of the systolic array and input/weight/output buffers, and the extensive workload demands of a neural network, 
tiling becomes necessary. We denote the tiled workload mapping using $L^0, C_{{in}}^0, C_{{out}}^0$ to represent the tile size. In CNNs, the output activation of one layer becomes the input activation of the subsequent layer. 
Thus, we set $C_{{in}}^0=C_{{out}}^0$ and $H=W$.

\textbf{Mapping.} \figref{hardware}~(center) illustrates the mapping of one matrix multiplication, corresponding to Lines 2-8 in~\figref{hardware} (right). The loops in Lines 3 and 6 are \textit{spatial-for}, while those in Lines 1-2 are \textit{temporary-for}. Concerning the weight, each $1\times1$ kernel, sized $(C_{{out}}^0, C_{{in}}^0)$, is spatially mapped to the height $H$ and width $W$ of the systolic array, respectively, and stored in the weight registers. For input activations with dimensions $(L^0, C_{{in}}^0)$, $L^0$ is temporarily mapped, while $C_{{in}}^0$ is spatially mapped to the width $W$. Similarly, the output activation has dimensions $(L^0, C_{out}^0)$, with $L^0$ temporarily mapped and $C_{{out}}^0$ spatially mapped to the height $H$. Each matrix multiplication yields partial results. To implement Lines 1 and 9 in~\figref{hardware} (right), the VPU initiates the addition of partial sums and updates the results. In each cycle, $C_{out}^0$ results form a vector and flow out of the systolic array in parallel. Simultaneously, the VPU retrieves the previous $C_{{out}}^0$ partial results from the output buffer using addresses provided by the address mapping scheme, adds the newly generated result to them, and writes the latest addition result back to the output buffer. Notably, an address detector precedes the addition to exclude contributions from certain edges that do not affect the output. Upon completion of the matrix multiplication related to the last $1\times1$ kernel, the partial results become the final result. In addition, the stride-$2$ convolution in \textit{StableDiff} is conveniently supported by changes in the input stride configuration from $1$  to $2$ or $W+2$ when spanning rows.

\textbf{Benefits.} The address-centric dataflow and mapping have the following advantages. \textit{1)} \textit{Memory regularity}: the dimension $l$ increases monotonically for both input and output, ensuring that bandwidth is efficiently utilized and equally distributed between them. \textit{2)} \textit{High PE utilization}: the mapping ensures high PE utilization for nearly all \textit{U-Net} layers, except for the first and last convolutions, which have small $C_{in}$ and $C_{out}$. However, their impact on latency is less than $1\%$, which does not affect overall performance. \textit{3)} \textit{Simplicity and generality}: convolution operations are efficiently managed by a naive systolic array, and the address mapping scheme can be easily implemented by adjusting the address bias of the output, with negligible overhead. \textit{4)} \textit{Compatibility}: the dimension $l$ of convolution aligns well with the sequence length dimension of attention, which avoids any format conversion overhead, making the unified support for convolution and attention highly effective. Thus, the remaining challenge is addressing the bottleneck of nonlinear operations.

\begin{figure*}[ht]
\centering
\includegraphics[width=180mm]{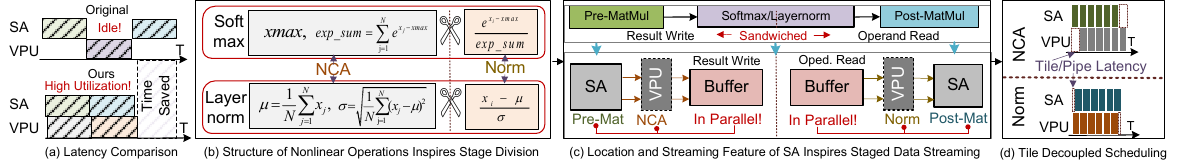}

\caption{\revision{(a) compares the latency of ours and the original, (b)-(d) shows three optimizations for dataflow scheduling. Notice that SA here contains the accumulation unit, because nonlinear operations are only performed on final results, and VPU specifically refers to arithmetic and function array of \figref{hardware}.}}
\label{schedule}
\vspace{-15pt}
\end{figure*} 
% \textbf{Support of  High-stride Convolution and Deconvolution.} \textit{StableDiff} employs stride-$2$ convolution for downsampling, while deconvolution achieves upsampling in the traditional \textit{U-Net}~\cite{ronneberger2015u}. Previously, these operators encountered bandwidth issues, and some accelerators designed dedicated engines for them~\cite{liu2018optimizing}. Fortunately, unified mapping can efficiently support them at no additional cost through the stride configuration of the input or output. For convolution with a stride of $2$, across each $1\times1$ kernel, the related input activation increases with a stride of $2$ or $W+2$ when spanning rows. A straightforward way to implement this is to change the stride configuration of the input buffer from $1$ to $2$ or $W+2$. For deconvolution, the stride of the output buffer is adjusted accordingly.
\subsection{Latency Hiding of Nonlinear Operations}\label{sec:nonlinear}
\textbf{Motivations.}
Nonlinear operations are indispensable for model inference, but they can even consume up to $30\%$ latency on GPU/CPU platforms \cite{kim2023stackoptimizationtransformerinference, spatten, 9586134, narechania2021vitality}. This issue is more challenging in edge devices due to limited hardware resources. The bottleneck of nonlinear operations, such as softmax and layernorm arises from two inefficiencies: \textit{(i)} They require multiple passes over the data (e.g., calculation of mean, variance, and normalization), with each pass introducing extra latency. Since they occur between Matmuls, this latency can block subsequent GEMV computations. \textit{(ii)} They require all data to be available before computation begins and prevent parallel computation until all required data is available. This stalled latency degrades the systolic array utilization, hence, we propose a \textit{2-stage streaming computing} strategy to address this issue, as illustrated in \figref{schedule} (a), where the comparison shows significant improvements.
% \begin{equation}
%   \textstyle
% softmax(x_{i})=\frac{e_{}^{x_{i}^{}}}{\sum_{j=1}^{N}e_{}^{x_{j}^{}}}=\frac{e_{}^{x_{i}-xmax}}{\sum_{j=1}^{N}e_{}^{x_{j}-xmax}}
% \label{softmax}
% \end{equation}
% \begin{equation}
%   \textstyle
% layernorm(x_{i})=\frac{{x_{i}}-\mu }{\sigma }=\frac{{x_{i}-\mu }}{\sqrt{\frac{1}{N}\sum_{j=1}^{N}(x_{j}-\mu)^{2}} }
% \label{layernorm}
% \end{equation}
% }

\revision{
\textbf{Observations and Optimizations.} Our approach builds on three key observations and optimizations. First, when evaluating the structure of these operations, softmax and layernorm can be split into two stages: numerical characteristic acquisition (NCA) and element-wise normalization (Norm), as illustrated in \figref{schedule} (b).
%for softmax, the maximum and exponential sum are computed during the NCA stage, while element-wise division is performed in the Norm stage. Similarly, for layernorm, the mean and variance are computed during the NCA stage, while subtraction and division occur in the Norm stage. %
Second, these operations occur between matrix multiplication (Matmul) operations.
Since systolic arrays generate results and read operands in a streaming fashion, it allows us to hide NCA and Norm latencies within the data streaming. Thus, as shown in \figref{schedule} (c), we schedule the NCA stage concurrently on the VPU with the preceding (pre-) matrix multiplication and insert the computation pipeline before the multiplication results are written to the buffer, enabling simultaneous NCA. Similarly, the Norm stage is scheduled on the VPU only whenever the posterior (post-) matrix multiplication is needed. The computation pipeline is inserted after the operand is read from the buffer, preparing normalization results concurrently. 
}

\textbf{Tile Decoupling.} These two steps address inefficiency-\textit{(i)}, while inefficiency-\textit{(ii)} is resolved by our mathematical transformation. Layernorm is transformed as shown in Equation (\ref{layernormtrans}), allowing the sum and square sum to be computed concurrently with the preceding matrix multiplication results, and the mean and variance can be calculated immediately after the matrix computation finishes. 
However, softmax requires a global maximum to avoid numerical overflow. Inspired by \cite{milakov2018onlinenormalizercalculationsoftmax}, we compute a partial exponential sum using the current maximum in advance, which is updated as new tiles are generated. As shown in Equation (\ref{expsum}), $ES$ represents the exponential partial sum of $N_1$ elements based on the previous global maximum ($pre\_max$), while $ES_n$ is the partial sum of $N_0$ elements from the new tile with the latest global maximum ($new\_max$). Each time a tile of size $N_0$ is generated, as shown in Equation (\ref{expsumtrans}), we update $ES$, increment $N_1$ by $N_0$, and simultaneously update the previous global maximum with $new\_max$. Based on these, our tile decoupled scheduling is shown in \figref{schedule} (d), where the VPU operates concurrently with generating tiled results from the SA during the NCA stage and prepares operand reads for the SA in the Norm stage. In the NCA stage of softmax, due to the tile-based maximum search, the VPU completes the computation with a slight delay of tile latency. For Norm, the VPU should be scheduled in advance to account for pipeline latency. For layernorm, both the NCA and Norm stages incur pipeline latency. It is worth noting that the only extra end-to-end latency is either tile or pipeline latency, both of which are negligible.

\textbf{Discussion.} Furthermore, compared with previous works employ a \textit{store-then-compute} strategy \cite{kim2023stackoptimizationtransformerinference,kao2023flat}, layernorm fusion requires all input data prepared and stored on the scratchpad, which restricts the Matmul tiling due to the limited scratchpad size, even possibly hurting the performance due to inefficient tiling \cite{kim2023stackoptimizationtransformerinference}. 
In contrast, with our \textit{2-stage streaming computing}, data can even move to off-chip, since their characteristics have been stored immediately in the NCA stage. Due to efficient streaming, nonlinear operations are decoupled from Matmul, imposing no restrictions on storage or tiling. 
\revision{
\begin{equation}
  \textstyle
\mu=\frac{1}{N}\sum_{j=1}^{N}x_{j};\sigma^2=\frac{1}{N}\sum_{j=1}^{N}x_{j}^{2}-(\frac{1}{N}\sum_{j=1}^{N}x_{j})^2
\label{layernormtrans}
\end{equation}
\begin{equation}
  \textstyle
ES=\sum_{j=1}^{N1}{e^{x_{j}-prev\_max}};ES_n=\sum_{j=1}^{N0}{e^{x_{j}-new\_max}}
\label{expsum}
\end{equation}
\begin{equation}
  \textstyle
ES\xleftarrow{}ES*e^{prev\_max-new\_max}+ES_n;N1\xleftarrow{}N1+N0
\label{expsumtrans}
\end{equation}
}
 \vspace{-10pt}
\begin{figure*}[ht]
\vspace{-10pt}
\centering
\includegraphics[width=180mm]{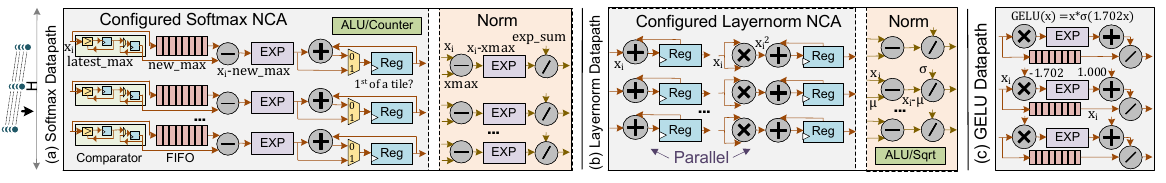}
\caption{\revision{Configured datapath for softmax, layernorm, and GELU by reconfigurable and reusable arithmetic and exponential arrays.}}
\label{vpu}
\vspace{-10pt}
\end{figure*}

\subsection{General Reconfigurable VPU for Nonlinear Operations}
\label{sec:VPU}
 \textbf{Motivations.} There is substantial work on designing specialized and efficient engines to approximately implement nonlinear operations \cite{yu2021nnlutneuralapproximationnonlinear,9586134,kim2021ibertintegeronlybertquantization}. However, this approach has limitations. First, while these circuits maintain accuracy for specific tasks, they may not perform as well in other scenarios. For example, I-GELU\cite{kim2021ibertintegeronlybertquantization} performs well with BERT\cite{devlin2019bertpretrainingdeepbidirectional} but shows a significant performance degradation for \textit{StableDiff} in our experiments. Second, dedicated engines are not flexible enough to support other operations, and designing multiple specialized engines for various operations can reduce overall efficiency. 

\textbf{Overview.} To uniformly support various nonlinear operations, we design a reconfigurable arithmetic and special function array of the VPU in~\figref{hardware}. This module comprises a comparator array, an exponential (EXP) array, a multiplier array, a divider array, two adder arrays, an ALU, and sufficient multiplexers. All arrays are $H-$parallel, with each row handling a single softmax or layernorm operation independently. The ALU is equipped with a register stack to store numerical characteristics. Since the entire input generates only two characteristics: exponential summation ($exp\_sum$) and global maximum ($xmax$) for softmax, mean and variance for layernorm, the storage cost is negligible.

\revision{
\textbf{Softmax Datapath.} As shown in \figref{vpu} (a) and Equation~(\ref{expsum}) (right), in the NCA stage with $H-$ parallel independent rows of data streams in, each input $x_{i}$ enters a comparator and FIFO, where the FIFO depth is consistent with the tile size (e.g., $32$). The comparator consists of two registers: the first continuously tracks the latest maximum, while the second is only updated by the latest maximum once a new tile has been entirely stored in the FIFO, and it represents the current maximum ($new\_max$) for exponential accumulation. Once the FIFO is fully filled, it outputs and receives data simultaneously. The output subtracts $new\_max$, enters the EXP unit, and is then accumulated. If the element is the first of a tile, the accumulation register is directly set as its value, eliminating the need for a clear cycle. The process is controlled under the ALU counter. Due to the naturally skewed data of the systolic array as shown in figure leftmost, ALU is reused row-by-row to implement cross-tile $exp\_sum$ updating, as described in Equation (\ref{expsumtrans}). In the Norm stage, each element subtracts the final maximum $xmax$, enters the reused EXP unit, and is divided by the final $exp\_sum$. \textbf{Layernorm Datapath.} As shown in \figref{vpu} (b), in the NCA stage based on Equation~(\ref{layernormtrans}), the sum and square sum are accumulated in parallel by two adder arrays, and the square sum computing requires the data multiplied by itself using the multiplier array. In the NCA stage, ALU conducts the mean and variance computing using the sum and square sum and performs the square root computation, which is again one-by-one reused among each row due to the skewed input for the systolic array. Finally, adder and division arrays are utilized for the normalization. In \textbf{GELU Datapath.} The official sigmoid version of GELU is implemented in our hardware \cite{hendrycks2016gaussian}, which has been validated to show negligible accuracy loss for \textit{StableDiff}. As shown in \figref{vpu} (c), the FIFO is reused to buffer and delay operands, while the EXP, adder, multiplier, and divider arrays are configured to reuse for computation.
}

\begin{figure*}[htb]
\vspace{-10pt}
\centering
\includegraphics[width=180mm]{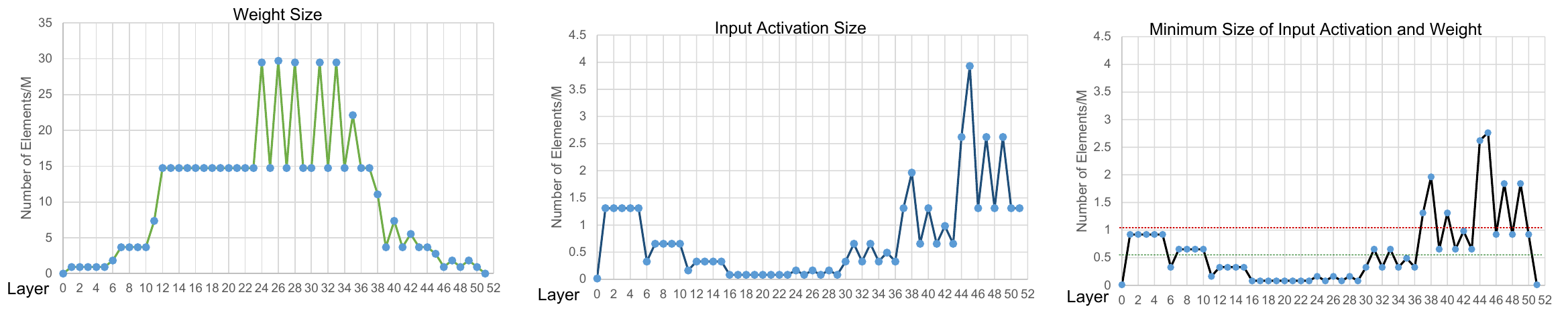}
\vspace{-8pt}
\caption{The variation of sizes across different convolutions in \textit{U-Net} of \textit{StableDiff} v $1.x$ and $2.x$.}
\label{variation}
\vspace{-15pt}
\end{figure*} 
\section{Adapative Dataflow Optimization}\label{sec:dataflow}
The \textit{U-Net} in \textit{StableDiff} integrates different upsampling and downsampling blocks, leading to substantial variations in weight and activation sizes across layers. This poses a challenge in maximizing data reuse and minimizing memory access, which will be discussed in~\secref{Varied Weight and Activation Size}. Therefore, we propose an adaptive dataflow optimization in~\secref {Adaptive Reuse and Fusion Methodology}.
\begin{figure*}[ht]
\centering
\includegraphics[width=180mm]{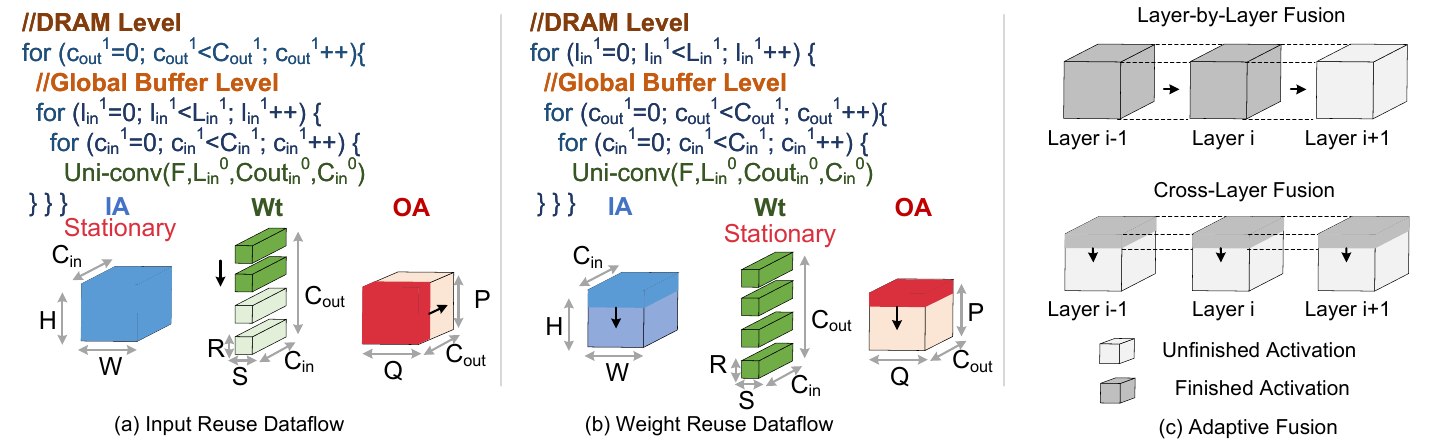}
\vspace{-5pt}
\caption{Adaptive input, weight reuse, with corresponding illustration (a), (b), and the adaptive fusion (c).}
\label{dataflow}
\vspace{-15pt}
\end{figure*} 
\subsection{Varied Weight and Activation Sizes}
\label{Varied Weight and Activation Size}
 \textbf{Observations.} The sizes of weight and input activation of $3\times3$ convolution across the entire \textit{U-Net} are shown in \figref{variation} (left) (center), indexed by $0-51$. Both activations and weights vary significantly, with the weight size nearly reaching 30M in certain layers. However, we observe an interesting phenomenon: in the shallowest and deepest layers, activations are large while weights are small, whereas in middle layers, activations are small but weights are large. Consequently, we plot the curve of the minimum size of input and weight across the layers, as shown in \figref{variation} (right). It can be concluded that: \textit{1)} For the minimum size, most layers are below a threshold, denoted by the red line. This inspires us to leverage the hybrid reuse methodology to minimize the off-chip memory traffic. \textit{2)} For middle layers, their minimum size is determined by the activation, and they are much smaller, as indicated by the green line. This inspires us to keep both the entire input and output on-chip and utilize fusion techniques for these layers.
\subsection{Adaptive Reuse and Fusion Methodology}
\label{Adaptive Reuse and Fusion Methodology}

Based on the address-centric dataflow, it comprises $(L_{in}, C_{in})\times(C_{in}, C_{out}) \xrightarrow{}(L_{in}, C_{out})$ matrix multiplications. Due to the limited global buffer size, tiling can be performed as $L_{in}=L_{in}^1*L_{in}^0$, where $L_{in}^0$ represents the tile size, $L_{in}^1$ denotes the number of tiles, and the tiling notation of $C_{out}$ and $C_{in}$ follows that. The design space is significantly simplified to a straightforward three-layer loop nest in \figref{dataflow}, and we employ the following strategies. \textbf{Considering a single layer:} For layers with small input activations and large weights, the input is small enough to be stored in the global buffer. We reuse it with the related weight tiles along the $C_{out}$ dimension, as shown in \figref{dataflow} (a), and \textit{Uni-conv} refers to~\figref{hardware} (right). For layers with large input activations and small weights, weight is reused in the global buffer and utilized with the related input tiles along the $L_{in}$ dimension, as shown in \figref{dataflow} (b). Through this adaptive reuse, for most layers, their input activations and weights can be accessed from off-chip only once. For the very few layers with both large weights and input activations, both of which exceed the global buffer, we can divide this layer into tiles. In summary, we consistently select the reuse method with less memory access for each layer. 

\textbf{Considering multiple layers:} As shown in \figref{dataflow} (c) (top), the input and output activations in the middle layers are relatively small enough to store both on-chip. Once one layer is fully processed, the activation is directly forwarded to the next layer, which is \textit{layer-by-layer fusion}. This approach can be applied to input or weight reuse, but it requires prioritizing buffer allocation for activations, which may lead to increased memory access of weight. For the shallowest and deepest layers, as shown in \figref{dataflow} (c) (bottom), the smaller weights make weight reuse feasible and enable the generation of streaming partial activations. Partial activations can be directly fed into the next layer’s execution before the entire activation is complete, eliminating off-chip access to intermediate activations, which is \textit{cross-layer fusion}. As this method requires the entire channel to be computed, it is only compatible with weight reuse. The weights of the fused layers must be stored on-chip simultaneously, which may exceed buffer capacity and result in more weight access, thus, the number of layers to be fused should be carefully selected. Particularly, after determining reuse, the fusion method should first be decided for input-reuse layers, as they can only utilize \textit{layer-by-layer fusion} or no fusion, followed by the decision for weight-reuse layers.

\section{Evaluation}
\subsection{Experiment Setup.}\label{subsec-exp-setup}
\textbf{Algorithm setup.} We select \textit{StableDiff} v$1.4$, v$2.1-base$, and \textit{StableDiff XL} as our targeted models because they are representative, influential, and open-source. For \textit{StableDiff} v$1.4$, v$2.1-base$, their latent resolutions are $64\times64$, which generates images with size $512\times512$. 
For \textit{StableDiff XL}, its latent resolution is $128\times128$, and the generated images are $1024\times1024$. The classifier-free guidance scale is set to the default value of $7.5$. We use the PNDM scheduler~\cite{liu2022pseudo} with a total of $50$ timesteps for sampling. We benchmark our method using the MS-COCO dataset validation split~\cite{lin2014microsoft}. This dataset contains human-generated captions for images. We randomly sample $5000$ captions from it as the prompt input for \textit{StableDiff}, thus generating $5000$ images with one prompt per image. Following popular settings~\cite{rombach2022high,ramesh2021zero}, each generated image is downsampled to $256\times256$ resolution for metrics evaluation. We employ the CLIP score with the ViT-g/14 model to evaluate the correspondence between the images and prompts, where higher scores indicate better performance. \revision{We compute the Fréchet Inception Distance (FID), indicating distributional differences between the generated images and the ground-truth real-world images, and the inception score (IS) to assess image quality.}

\textbf{Hardware implementation.} We utilize SystemVerilog and Vivado 2018.3 to synthesize and implement the hardware modules using the VCU118 Xilinx FPGA. The systolic array has a size of $32\times32$, the vector processing unit is $32-$ parallel, and the arithmetic for the hardware is fp16. The hardware details are listed in Table \ref{tb: hardware}. Our energy consumption contains the on-chip cost and off-chip access, and the latter is derived from the access behavior and based on ~\cite{energy}. We also adopt a cycle-accurate performance model to measure latency and memory traffic. 
\begin{table}[hbtp]
\vspace{-10pt}
\centering
\caption{Resource and power consumption of our accelerator.}
\label{tb: hardware}
\scalebox{0.95}{
\begin{tabular}{cccccc}
\hline
\multicolumn{1}{c|}{Module/Resources} & LUT     & FF      & DSP  & \multicolumn{1}{c|}{BRAM} & Power \\ \hline
\multicolumn{1}{c|}{Available}        & 1182240 & 2364480 & 6840 & \multicolumn{1}{c|}{2160} & /     \\
\multicolumn{1}{c|}{Systolic Array}   & 405760  & 52672   & 0    & \multicolumn{1}{c|}{0}    & 11.30W \\
\multicolumn{1}{c|}{Vector Processing Unit}     & 40076 & 13798 & 288 & \multicolumn{1}{c|}{0}    & 0.98W  \\
\multicolumn{1}{c|}{Global Buffer}    & 1028    & 472     & 0    & \multicolumn{1}{c|}{456}  & 0.91W \\
\multicolumn{1}{c|}{Input, Weight, Output Buffer} & 628   & 166   & 0   & \multicolumn{1}{c|}{58} & 0.14W \\ \hline
\multicolumn{6}{c}{Total Power = 15.98 W,    Frequency = 200MHz,   DDR = 38.4GB/s}                   \\ \hline
\end{tabular}}
\vspace{-5pt}
\end{table}

\textbf{Baselines.} For algorithm evaluation, we conduct the comparison with the original $50$ timesteps original model. To compare with the state-of-the-art works, Deepcahe~\cite{ma2023deepcache}
and BK-SDM~\cite{kim2023bk} are selected because they are excellent open-sourced designs, optimizing the \textit{U-Net} for \textit{StableDiff} as well. Deepcache also utilizes layer skipping and BK-SDM employs the knowledge distillation. For hardware evaluation, we compare our design with the AMD Ryzen 7 6800H CPU, Intel(R) Xeon(R) Gold 5220R CPU, and NVIDIA V100 GPU. Additionally, we also conduct the ablation study to evaluate the gain of the address-centric adaptive dataflow optimization.
\subsection{Algorithm Evaluation}
\label{sec: alg_eval}

\begin{table}[hbtp]
\vspace{-10pt}
\centering
\caption{\revision{The image quality and performance gain of our approach under different configurations.}}
\label{tb: compression_config}
\scalebox{0.79}{\revision{\begin{tabular}{c|cccc|cccc|ccc}
\hline
         & \multicolumn{4}{c|}{StableDiff 1.4} & \multicolumn{4}{c|}{StableDiff 2.1} & \multicolumn{3}{c}{StableDiff XL} \\ \hline
\begin{tabular}[c]{@{}c@{}}Model/\\ Metric\end{tabular} &
  \begin{tabular}[c]{@{}c@{}}CLIP\\ Score↑\end{tabular} &
  \begin{tabular}[c]{@{}c@{}}FID\\ ↓\end{tabular} &
  \begin{tabular}[c]{@{}c@{}}IS\\ ↑\end{tabular} &
  \begin{tabular}[c]{@{}c@{}}MAC \\ Red.↑\end{tabular} &
  \begin{tabular}[c]{@{}c@{}}CLIP\\ Score↑\end{tabular} &
  \begin{tabular}[c]{@{}c@{}}IS\\ ↑\end{tabular} &
  \begin{tabular}[c]{@{}c@{}}FID\\ ↓\end{tabular} &
  \begin{tabular}[c]{@{}c@{}}MAC \\ Red.↑\end{tabular} &
  \begin{tabular}[c]{@{}c@{}}CLIP\\ Score↑\end{tabular} &
  \begin{tabular}[c]{@{}c@{}}FID\\ ↓\end{tabular} &
  \begin{tabular}[c]{@{}c@{}}MAC \\ Red.↑\end{tabular} \\ \hline
Original & 0.3004   & 25.38   & 32.9   & 1.00  & 0.3103   & 31.72   & 19.33  & 1.00  & 0.3230     & 17.32     & 1.00     \\
PAS-25/2 & 0.2989   & 24.53   & 32.21  & 2.39  & 0.3109   & 32.22   & 18.88  & 2.49  & 0.3207     & 18.24     & 3.22     \\
PAS-25/3 & 0.2982   & 24.25   & 31.64  & 2.72  & 0.3105   & 32.47   & 18.35  & 2.84  & 0.3198     & 18.18     & 3.96     \\
PAS-25/4 & 0.2978   & 24.01   & 31.63  & 2.84  & 0.3096   & 31.79   & 18.00  & 2.98  & 0.3191     & 17.88     & 4.28     \\
PAS-25/5 & 0.2966   & 23.79   & 31.42  & 3.31  & 0.3088   & 31.60    & 17.61  & 3.50  & 0.3177     & 18.12     & 5.68     \\ \hline
\end{tabular}}}
%\vspace{-5pt}
\end{table}
\textbf{Evaluation of various configurations.} Our phase-aware sampling (PAS) is adequately evaluated with various configurations. As shown in Table \ref{tb: compression_config}, the original model requires $50$ timesteps execution of the complete \textit{U-Net}. In our phase-aware sampling, we change the $T_{sparse}$ and fix other hyper-parameters to test the image quality and MAC reduction. Across  three different models, $T_{sketch}$ is set as $25$, as denoted by '$-25$', and $L_{sketch}=L_{refine}=2$. For \textit{StableDiff} v$1.4$, $T_{complete}$ is set as $4$, while for the others $T_{complete}=3$. $T_{sparse}$ is represented by '$/*$'.
For instance, $/4$ represents executing the complete \textit{U-Net} once out of every four timesteps in the sketching phase. Our method achieves explicit MAC reduction for all the models. For image quality, we achieve CLIP scores similar to those of the original model. \revision{Interestingly, we achieve better FID and IS scores in some configurations. }
It may be because our method incorporates prior human knowledge, which has a regularization effect. This prevents the model from overfitting to the training datasets, making generated images closer to the real world. The effectiveness of the experimental result further demonstrates the $phase-divison$ phenomenon. As the MAC reduction increases, the CLIP score decreases and degrades the image quality, indicating the upper bound of MAC reduction. 
% \begin{figure*}[ht]
% \centering
% \includegraphics[width=170mm]{./figs/access_breakdown.pdf}
% \vspace{-10pt}
% \caption{Off-chip access reduction across the convolution layers of \textit{StableDiff} by adaptive reuse and fusion.}
% \label{access}
% \vspace{-10pt}
% \end{figure*} 

\textbf{Comparison with the state-of-the-art work.} We select \textit{StableDiff} v$1.4$ for comparison. As shown in Table \ref{tb: compression_sota}, our method outperforms BK-SDM and Deepcache in both inference performance and image quality. BK-SDM \cite{kim2023bk} compresses the \textit{U-Net} by employing block pruning and feature distillation of the \textit{U-Net}, resulting in MAC reduction and structured network architecture. However, their knowledge distillation requires retraining and may lead to overfitting, as indicated by the poor FID. Deepcache \cite{ma2023deepcache} adopts layer-skipping uniformly, but it does not realize the mechanism of phase division. In contrast, our method utilizes phase division to perform an adaptive sampling scheme without the need for retraining. Our general framework also allows for exploring the trade-off between image quality and acceleration. 
% which is compatible with fewer original timesteps and requires no retraining.
\begin{table}[hbtp]
\vspace{-10pt}
\centering
\caption{Comparison with state-of-the-art approaches.}
\label{tb: compression_sota}
\scalebox{0.95}{
\begin{tabular}{ccccc}
\hline
Model/Metric & CLIP Score↑ & FID↓  & MAC Red.↑ & Speedup (GPU)↑ \\ \hline
Original     & 0.3004      & 25.38 & 1              & 1               \\ \hline
BK-SDM-Base  & 0.2919      & 29.16 & 1.51           & 1.46            \\
BK-SDM-Small & 0.2713      & 31.77 & 1.56           & 1.57            \\
BK-SDM-Tiny  & 0.2684      & 31.74 & 1.65           & 1.79            \\
Deepcache    & 0.2980      & 24.54 & 2.11           & 1.66            \\
PAS-25/4     & 0.2978      & 24.01 & \textbf{2.84}  & \textbf{2.45}   \\ \hline
\end{tabular}}
\vspace{-10pt}
\end{table}

\subsection{\revise{Hardware Ablation Study}}
\label{sec:hardware-ab}
\revise{In this subsection, we focus on the FPGA inference of the original dense model, and select \textit{StableDiff} v$1.4$ as a typical case to conduct an ablation study for each hardware optimization.}
\begin{figure}[htbp!]
\vspace{-5pt}
\centering
\includegraphics[width=88mm]{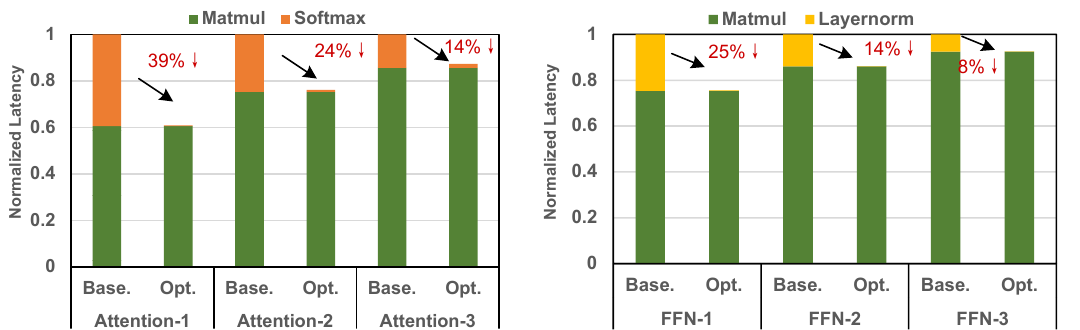}
\vspace{-10pt}
\caption{\revise{Latency reduction by streaming computing.}}
\label{stream-ab}
\vspace{-10pt}
\end{figure} 
\begin{figure}[h]
\vspace{-10pt}
\centering
\includegraphics[width=88mm]{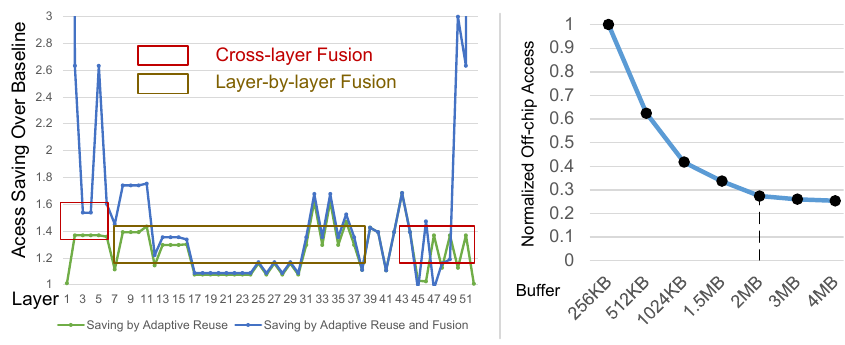}
\vspace{-10pt}
\caption{The performance gain brought by adaptive fusion (left) and the exploration of the global buffer size (right).}
\label{fusion}
\end{figure} 

\begin{figure*}[t]
\vspace{-10pt}
\centering
\includegraphics[width=180mm]{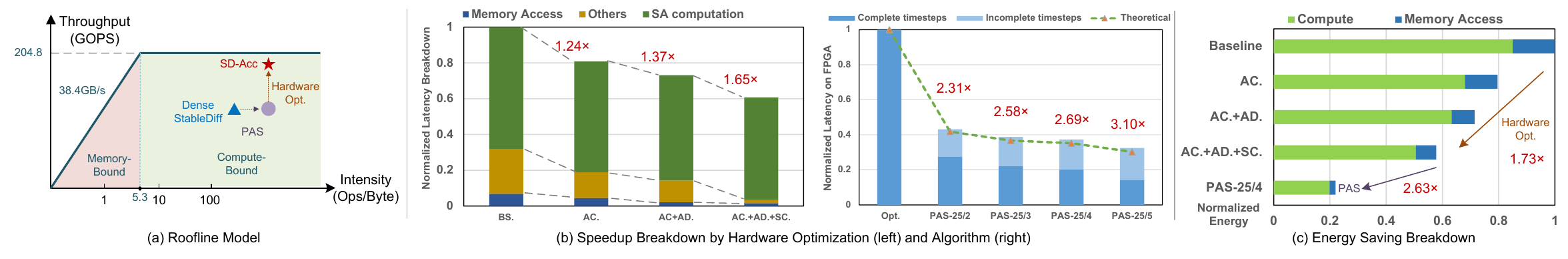}
\vspace{-5pt}
\caption{ \revise{Algorithmic and hardware optimizations breakdown analysis.}}
\label{tech-break}
\vspace{-15pt}
\end{figure*} 

\revise{\textbf{2-stage Streaming Computing Study.} Since the latency of nonlinear operations is related to the sequence length and the hidden dimension of the Transformer, we extract different Transformer layers to study the effect of 2-stage streaming computing, with self-attention and FFN analyzed separately. These layers are labeled as $-1$, $-2$, and $-3$, with sequence lengths (i.e., feature map resolutions) of $4096$, $1024$, and $256$, respectively. The baseline shares the same hardware settings as the optimized solution, except for the use of 2-stage streaming computing. As demonstrated in \figref{stream-ab} (left) for self-attention, our optimization nearly eliminates the entire latency of softmax, achieving latency reductions of $39\%$, $24\%$, and $14\%$, respectively, compared with the baseline. The effect is more pronounced for larger sequence lengths, as the softmax operations scale quadratically. For FFN, as shown in \figref{stream-ab} (right), our solution achieves latency reductions of $25\%$, $14\%$, and $8\%$. The reduction is smaller because compared to the heavy matrix multiplications of the FFN, layernorm occupies a smaller portion of the computation. The savings are due to 2-stage streaming computing, on one hand, it avoids multiple passes of the data, improving memory efficiency, and on the other hand, absorbs computation into the data streaming of the systolic array's read and write operations. Additionally, it decouples data dependencies and eliminates systolic array idleness caused by bottlenecks from nonlinear operations.
}

\textbf{Adaptive Dataflow Optimization Study.} Various operators are uniformly supported by the systolic array through address-centric dataflow, and our adaptive reuse and fusion optimization can further reduce off-chip access. \revise{To analyze the effect of this technique, we set the baseline with a systolic array of the same size, equipped with an \textit{im2col} hardware module, following the design in ~\cite{genc2021gemmini, jouppi2017datacenter}. The baseline is also equipped with streaming computing for nonlinear operations, and it has a global buffer of $2$MB and a memory bandwidth of $38.4$ GB/s, identical to the optimized solution for fairness.} In our dataflow optimization, we dynamically employ the most suitable reuse scheme with less access between input and weight reuse, adapting to the varying weight and activation ratio across convolution layers. 
% As demonstrated in \figref{access}, adaptive reuse consistently achieves reduced access, and adaptive further reduces activation access by considering multiple layers. 
For most layers, input, weight, and output are accessed from off-chip only once by adaptive reuse, and fusion also avoids repeated input and output activation access. Considering all layers, adaptive reuse and fusion save $24.3\%$ and $30.5\%$ of off-chip access in total, respectively.

In more detail, as shown in \figref{fusion} (left), cross-layer fusion is employed for layers $0\sim5$ and $44\sim51$. Because for these layers, the activation size is large and the weight is small, we can reuse the weight on-chip and fuse the activations part by part, as mentioned in \figref{dataflow} (c). Due to the large activation sizes, the savings are evident. Layer-by-layer fusion is employed for layers $6\sim36$ because the input and output activations of these layers are small and weights are large, making storing both activations in the global buffer feasible. However, due to the large weight size proportion, the reduction ratio is relatively smaller.
For other layers, no fusion is adopted because we find that the increment of weight access surpasses the reduction of activation. Through this adaptive reuse and fusion, we not only reduce power consumption by minimizing memory access but also save bandwidth to enhance performance. Additionally, utilizing the optimization, the influence of global buffer size is explored as shown in \figref{fusion} (right), and the off-chip access with the global buffer of $256$KB is normalized to one. The global buffer of $2$MB is observed to achieve the sweet spot, while a larger buffer leads to higher costs and lower buffer utilization.

\revise{\subsection{Technique Breakdown Analysis}}
\label{sec:tech-break}
\revise{In this subsection, we put the algorithmic and hardware optimizations together on the FPGA platform to analyze the speedup and energy efficiency improvement of each technique for \textit{StableDiff} v$1.4$ as a typical case. 
}

\revise{
\textbf{Roofline Model Analysis.} As shown in \figref{tech-break} (a), the roofline model indicates that the inference of the \textit{StableDiff} series on our FPGA platform (with a peak throughput of $204.8$ GFLOPS and a bandwidth of $38.4$ GB/s) is compute-bound. \textit{Through phase-aware sampling}, operational intensity increases because the timesteps executing only the topmost \textit{U-Net} blocks are performed on the large resolution, providing more opportunities for convolution kernel reuse compared to the blocks below. Similarly, the Transformers in these top blocks also have longer sequence lengths. As a result, with more aggressive compression (larger $T_{sparse}$), the averaged workload becomes increasingly compute-bound. \textit{Through our hardware optimizations}, our solution significantly improves real throughput, bringing it close to the theoretical maximum.}

\revise{\textbf{Speedup Analysis.} The effects of hardware and algorithmic optimizations are shown in \figref{tech-break} (b), left and right. For hardware optimization, we break down latency into memory access, systolic array computation, and others, which include nonlinear operations, \textit{im2col} operations, and other minor latencies. The baseline consists of a systolic array equipped with an \textit{im2col} module, following the design in ~\cite{genc2021gemmini}, and the latency is normalized to $1$. Our hardware optimizations include address-centric dataflow (AC.), adaptive dataflow optimization (AD.), and 2-stage streaming computing (SC.). As shown in \figref{tech-break} (b) left, compared to the \textit{im2col} implementation, the address-centric dataflow implementation achieves a $1.24\times$ speedup. This speedup is due to the original \textit{im2col} operation causing irregular memory access and potential bank conflicts, whereas address-centric dataflow maintains regular memory access and effectively utilizes bandwidth. Furthermore, the \textit{im2col} module introduces explicit latency, which can bottleneck systolic array computation, especially when the kernel, feature map size, and convolution stride vary. After applying adaptive dataflow optimization, our solution achieves a $1.37\times$ speedup over the baseline, due to reduced memory access and improved systolic array PE utilization. Finally, 2-stage streaming computing effectively hides the latency introduced by nonlinear operations, achieving a $1.65\times$ speedup. Taking the fully optimized hardware as the baseline, phase-aware sampling, depending on $T_{sparse}$, achieves a speedup ranging from $2.31\times$, $2.58\times$, $2.69\times$ and $3.10\times$, respectively, as shown in \figref{tech-break} (b) right. With a larger $T_{sketch}$, more timesteps replace the incomplete \textit{U-Net} with the complete one, and our hardware implementation achieves nearly $95\%$ of the theoretical speedup due to the highly optimized hardware implementation.}

\revise{\textbf{Energy Saving Analysis.} As shown in \figref{tech-break} (c), compared to the baseline implementation, hardware optimizations achieve a $1.73\times$ reduction in energy consumption due to reduced latency and memory access. Additionally, phase-aware sampling further achieves a $2.63\times$ energy savings. It is important to note that on-chip computation energy still dominates consumption, as an FPGA consumes over $5\times$ more power than an ASIC implementation at the same process node \cite{kuon2006measuring}. Therefore, the energy savings would be more significant in an ASIC implementation, which we plan to explore in future work.}

\subsection{\revise{Comparison with SOTA Accelerators}}
\label{sec:sota-comp}
\begin{figure}[h]
\vspace{-12pt}
\centering
\includegraphics[width=66mm]{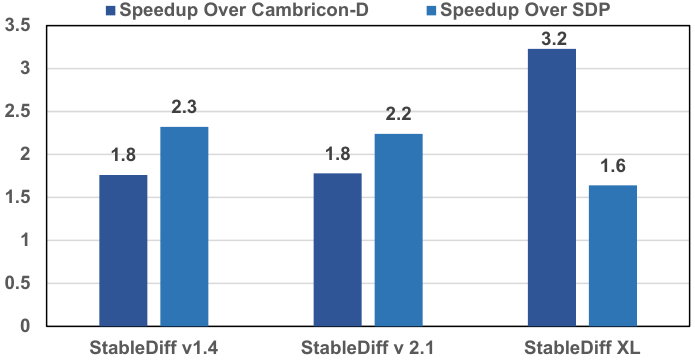}
\vspace{-5pt}
\caption{\revise{Speedup over SOTA  \textit{StableDiff} accelerators.}}
\label{sota-comp}
\vspace{-10pt}
\end{figure} 
\revise{We select the SOTA \textit{StableDiff} accelerators, Cambricon-D \cite{cambriconD2024isca} and SDP \cite{10558026}, for comparison. We built simulators based on the details provided in their papers to measure latency. Cambricon-D employs differential computing to optimize convolution layers by leveraging feature map similarity between consecutive timesteps. SDP utilizes prompt information to identify unimportant tokens through cross-attention computation, thereby accelerating the following FFN computation. Since Cambricon-D has the highest peak throughput, both SDP and our \textit{SD-Acc} are scaled to the same peak throughput. Memory bandwidths are set to that of Cambricon-D. As shown in \figref{sota-comp}, with PAS-25/4, \textit{SD-Acc} achieves a $1.8\times$ to $3.2\times$ speedup over Cambricon-D and a $1.6\times$ to $2.3\times$ speedup over SDP across three \textit{StableDiff} models. This difference arises from the varying ratios of convolutions to Transformers. Transformers occupy a larger proportion in \textit{StableDiff XL}, reducing Cambricon-D's acceleration effect, while the acceleration of SDP becomes more pronounced. In contrast, our \textit{SD-Acc} performs consistently well due to algorithm and hardware co-design, at the algorithmic level, phase-aware sampling skips ineffective computations in both convolution and Transformer layers, and at the hardware level, our optimizations effectively enhance the PE utilization and optimize the memory access, which puts theoretical gain into practice.}

\subsection{Comparison with CPU and GPU} The configurations shown in \secref{sec: alg_eval} are adopted to conduct the comparison with CPU and GPU.

\textbf{Energy Comparison.} As shown in \figref{energy}, comparing with the original model on AMD 6800H CPU ($6$nm), Intel 5220R CPU ($14$nm), and NVIDIA V100 GPU ($12$nm), our VCU118 FPGA implementation (\revise{$16$nm is the least advanced process among them.}) \revise{still} achieves $14.7\sim37.3\times$, $18.3\sim44.9\times$, $2.7\sim6.0\times$ energy saving across three models. The saving mainly comes from two aspects. First, phase-aware sampling reduces both computational and memory overhead. Second, our systematic hardware optimizations effectively reduce latency, and adaptive reuse and fusion decrease off-chip accesses.  It is worth noting that we choose the NVIDIA V100 GPU since it belongs to the same generation as the VCU118, \revise{and our solution is expected to achieve even greater energy savings with the ASIC implementation \cite{kuon2006measuring}.} %This choice also follows the practices described in~\cite{you2023vitcod,narechania2021vitality}.
\begin{figure}[hbtp]
\vspace{-15pt}
\centering
\includegraphics[width=88mm]{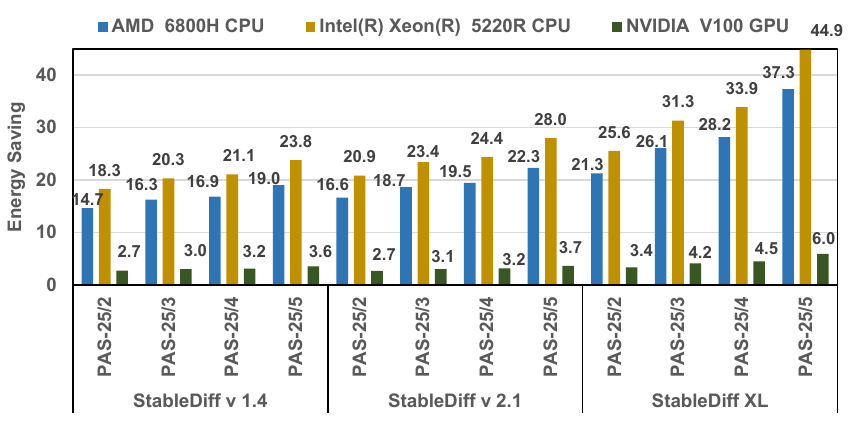}
\vspace{-10pt}
\caption{Energy saving of our approach over baselines.}
\label{energy}
\vspace{-10pt}
\end{figure} 

\begin{figure}[htp]
\centering
\vspace{-10pt}
\includegraphics[width=88mm]{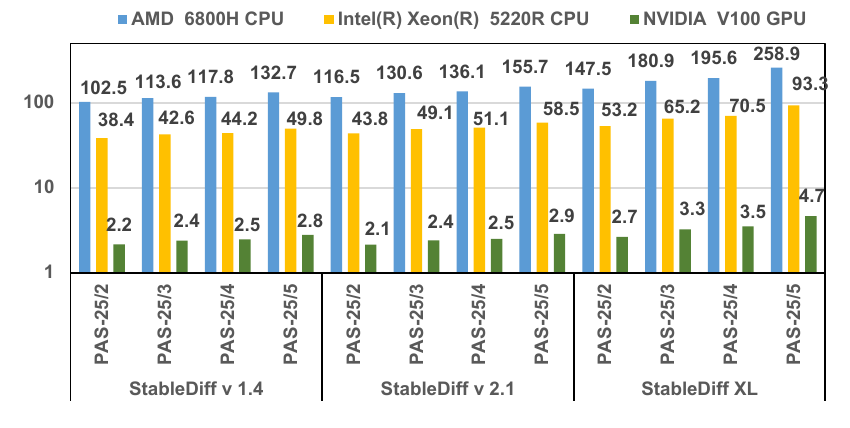}
\vspace{-15pt}
\caption{Scaled speedup of our approach over baselines.}
\label{speedup}
\vspace{-10pt}
\end{figure} 

\textbf{Speed Comparison.} The NVIDIA V100 GPU has a peak throughput of $14$ TFLOPS, while our FPGA-based implementation only possesses a peak throughput of $0.2$ TFLOPS. To make a fair comparison and keep consistent with previous accelerators~\cite{lu2021sanger,narechania2021vitality}, we scale the frequency of the accelerator from $200$MHz to $1$GHz and the MAC unit count from $1024$ to $4096$. As depicted in \figref{speedup}, for various configured three models, we achieve speedups ranging from $102.5\sim258.9\times$, $38.4\sim93.3\times$, and $2.2\sim4.7\times$ over the original model on AMD 6800H CPU, Intel 5220R CPU, and NVIDIA V100 GPU respectively. These speedups are achieved through algorithmic improvement, dataflow optimization, and dedicated hardware design. For the algorithm, our phase-aware sampling effectively reduces the operations, for hardware and dataflow, the address-centric dataflow and 2-stage streaming computing make the systolic array uniformly and efficiently support both convolution and attention, achieving high parallelism and hardware utilization.

\section{Related Work}\label{sec:related_work}
\textbf{Diffusion Algorithm Optimizations.} To accelerate the inference of diffusion models, recent efforts can be mainly classified into \textit{i)} reducing denoising timesteps and \textit{ii)} optimizing the architecture of \textit{U-Net}. 
For timestep reduction, various approaches such as fast solvers, distilled models, and early stopping have been explored~\cite{lu2022dpm,sauer2024fast,meng2023distillation,sauer2023adversarial,lyu2022accelerating}. 
In terms of efficient architecture, 
techniques such as pruning, quantization, and knowledge distillation have been employed to enhance \textit{U-Net}~\cite{fang2024structural,ma2023deepcache,li2023q,kim2023bk,li2024snapfusion}. 
However, these methods don't realize the phenomenon of phase division and overlook inference efficiency on the underlying hardware.

\revise{\textbf{Phase-aware Sampling.}  Recently, some studies also found that the importance of each timestep varies throughout the denoising process in diffusion models \cite{yang2023denoising, pan2024t}. For example, \cite{yang2023denoising} trains a slimmable network from scratch to replace the conventional \textit{U-Net}, allowing the network size to adapt at each denoising timestep. Building on this, the authors propose an evolutionary algorithm to search for the optimal step-aware strategy among various candidates, suggesting that each timestep may execute a denoising network of varying sizes. However, the training and search process takes $3\times$ longer than the original training, resulting in expensive computational overhead. Similar to this idea, \cite{pan2024t} proposes using different sizes of denoising networks for different timesteps. However, acquiring different-sized denoising networks still requires training, fine-tuning, or distillation. Moreover, these studies are based on empirical experiments but fail to uncover the underlying mechanisms. Our work uniquely demonstrates that different levels of neural network features play varying roles throughout the denoising process, revealing the significant phase-division phenomenon and showing that denoising from scratch is more computationally intensive than denoising from a sketch. Based on these observations, we propose phase-aware sampling, which can be seamlessly applied to the original \textit{U-Net}, without requiring additional candidate networks, and avoids retraining or fine-tuning. By accurately targeting the source of redundancy, our solution achieves the most significant performance improvement at the lowest cost compared to these related works.}

\textbf{Hardware Accelerators.} Various domain-specific accelerators have been designed for CNNs ~\cite{chen2016eyeriss,chen2019eyeriss,chen2014diannao,zhang2015optimizing,alwani2016fused,judd2016stripes,nvdla} and Transformers ~\cite{spatten,a3, dota,lu2021sanger,you2023vitcod, butter}. However, while effective within their domains, Transformer accelerators are not well-suited for executing convolutional operations, and vice versa for CNN accelerators when applied to Transformers. Since \textit{StableDiff} uniquely combines two distinct types of neural networks as cascaded blocks, the integration of two separate engines may result in hardware underutilization. \revise{To enable the efficient deployment of complex models on edge devices, exemplified by \textit{StableDiff}, we propose effective and broadly applicable hardware optimizations, including address-centric dataflow, 2-stage streaming computing, and adaptive dataflow optimizations.}
% Recently, Cambricon-D~\cite{cambriconD2024isca} introduced a differential computing architecture to accelerate guided diffusion, which mainly consists of convolutions. However, It necessitates an extremely large on-chip buffer to reduce the heavy off-chip access, and the algorithmic feasibility for \textit{StableDiff} remains to be explored. Our work addresses the hardware-level challenges of accelerating \textit{StableDiff} by proposing address-centric dataflow, 2-stage streaming computing, and adaptive dataflow optimizations.

\revise{\textbf{Unified Systolic Array for Linear Operators.}  Systolic arrays can effectively support matrix multiplications. Several works extend the original architecture with additional buses, buffers, routers, and control logic to efficiently support convolutions \cite{chen2016eyeriss, chen2019eyeriss, lu2017flexflow}. However, such extensions often introduce significant overhead, degrading both area and energy efficiency for basic matrix multiplication. Another common solution is to use \textit{im2col} to transform convolution into matrix multiplication, which has seen successful deployment in both industry \cite{liao2019davinci, jouppi2017datacenter} and academia \cite{genc2021gemmini, kung2019packing, 10.1109/ISCA45697.2020.00086, wei2017automated}. However, as reported in \cite{genc2021gemmini, soltaniyeh2022accelerator}, \textit{im2col} in software can account for up to $30\%$ of the end-to-end latency and result in a significant increase in memory access. Placing a dedicated \textit{im2col} hardware module can alleviate this issue. However, it still faces bank conflicts due to irregular memory access \cite{soltaniyeh2022accelerator}. Additionally, frequent conversions between formats introduce explicit latency, which is further aggravated by varying feature map shapes, kernel sizes, and strides. In contrast, our address-centric dataflow offers a competitive alternative. It addresses memory irregularities and enhances PE utilization without modifying the systolic array hardware architecture, making it a nearly cost-free and effective plug-in for conventional systolic arrays.}

\revise{\textbf{Fusion of Nonlinear Operations.} The fusion of complex nonlinear operations, such as softmax and layernorm, of Transformers has been studied in \cite{kao2023flat, kim2023stackoptimizationtransformerinference, dao2022flashattentionfastmemoryefficientexact, dao2023flashattention2fasterattentionbetter}. While these works propose efficient solutions to reduce memory access and enhance the operational intensity of matrix multiplication, they do not adequately address the inherent latency of nonlinear operators. For instance, \cite{kao2023flat} does not account for the latency of softmax. As reported in \cite{hong2024flashdecodingfasterlargelanguage}, the softmax update operation (not complete softmax) can still consume $18.8\%$ of the latency in attention computation in Flash-Attention. Moreover, \cite{kim2023stackoptimizationtransformerinference} reports that solely pursuing fusion may degrade performance, because limited buffer size could hinder the efficient tiling strategy. In contrast, our 2-stage streaming computing is proposed to solve these deficiencies systematically. It is both compatible with and orthogonal to the dataflow optimization of Flash-Attention, and it further addresses the latency issue comprehensively. It divides the nonlinear operations into two stages and integrates them into the mandatory systolic array write and read process, eliminating the need for multiple data passes. By registering only features instead of the entire data, it eliminates the need for a large buffer size. Finally, it decouples the data dependency and hides the latency in the data streaming process of the systolic array. Moreover, our reconfigurable vector processing unit uniformly supports various nonlinear operations with a single engine design without sacrificing generality or accuracy, as seen in previous approximation solutions \cite{yu2021nnlutneuralapproximationnonlinear, 9586134, kim2021ibertintegeronlybertquantization}.}

\textbf{Dataflow Flexible Accelerators.} Existing dataflow flexible accelerators typically rely on reconfigurable on-chip networks for dataflow switching, such as \cite{kwon2018maeri,qin2020sigma,tong2024feather}. 
% MAERI \cite{kwon2018maeri} introduces tree-based interconnects, uniformly supporting both CNN and Matmul. SIGMA \cite{qin2020sigma} introduces a forwarding adder network to achieve training acceleration. FEATHER \cite{tong2024feather} introduces a reconfigurable accelerator capable of flexible data layout reordering. 
Despite their effectiveness, the flexibility incurs high hardware overhead and bandwidth demands. \textit{SD-Acc} proposes a compromise solution: At the SA-to-buffer level, it adopts a fixed address-centric dataflow to unify Convolution and Matmul computations, minimizing the overhead increment of SA. At the buffer-to-memory level, specifically tailored for \textit{StableDiff}, it employs adaptive dataflow optimization to reduce off-chip access, which is a major contributor to energy consumption \cite{sze2020efficient}.

\section{Conclusion}\label{sec:conlcusion}

This paper proposes a novel approach to accelerate \textit{StableDiff} via algorithm and hardware co-optimizations.
By investigating the statistics of \textit{StableDiff},
this work proposes a novel phase-aware sampling, which adaptively \revision{applies approxiate} computation through an optimization framework without harming the image quality.
To accelerate the models with multiple distinct operators,
we propose an address-centric dataflow, 2-stage streaming computing, and adaptive dataflow optimization, significantly improving the hardware performance and efficiency.
Combining both algorithm and hardware optimizations,
our solution achieves $2.7\sim6.0\times$ higher energy efficiency than the NVIDIA GPU implementation.

% \textcolor{red}{Place holder:}
% The increasing demand for running multiple DNNs in parallel and the prevalence of sparsity across different DNN models have led to the emergence of sparse multi-DNN workloads.
% By identifying the optimization opportunities in sparse multi-DNN workloads,
% we propose a novel bi-level dynamic and static scheduler 
% that utilizes sparsity dynamicity and pattern information for better scheduling.
% Coupled with an efficient hardware scheduler and sparse latency predictor,
% our proposed approach achieves up to $10$\% fewer violations and nearly $4\times$ lower average normalized turnaround time compared to the state-of-the-art methods while incurring negligible hardware cost. 
% To facilitate future development in this area,
% we will open-source all the benchmarks and code upon the paper's acceptance.
% We believe our contributions will attract further research attention to study sparse multi-DNN workloads.

%%%%%%% -- PAPER CONTENT ENDS -- %%%%%%%%

%%%%%%%%% -- BIB STYLE AND FILE -- %%%%%%%%
\bibliographystyle{IEEEtranS}
\bibliography{refs}

% Generated by IEEEtranS.bst, version: 1.13 (2008/09/30)
\begin{thebibliography}{10}
\providecommand{\url}[1]{#1}
\csname url@samestyle\endcsname
\providecommand{\newblock}{\relax}
\providecommand{\bibinfo}[2]{#2}
\providecommand{\BIBentrySTDinterwordspacing}{\spaceskip=0pt\relax}
\providecommand{\BIBentryALTinterwordstretchfactor}{4}
\providecommand{\BIBentryALTinterwordspacing}{\spaceskip=\fontdimen2\font plus
\BIBentryALTinterwordstretchfactor\fontdimen3\font minus
  \fontdimen4\font\relax}
\providecommand{\BIBforeignlanguage}[2]{{%
\expandafter\ifx\csname l@#1\endcsname\relax
\typeout{** WARNING: IEEEtranS.bst: No hyphenation pattern has been}%
\typeout{** loaded for the language `#1'. Using the pattern for}%
\typeout{** the default language instead.}%
\else
\language=\csname l@#1\endcsname
\fi
#2}}
\providecommand{\BIBdecl}{\relax}
\BIBdecl

\bibitem{alwani2016fused}
M.~Alwani, H.~Chen, M.~Ferdman, and P.~Milder, ``Fused-layer cnn
  accelerators,'' in \emph{2016 49th Annual IEEE/ACM International Symposium on
  Microarchitecture}.\hskip 1em plus 0.5em minus 0.4em\relax IEEE, 2016, pp.
  1--12.

\bibitem{chen2014diannao}
T.~Chen, Z.~Du, N.~Sun, J.~Wang, C.~Wu, Y.~Chen, and O.~Temam, ``Diannao: A
  small-footprint high-throughput accelerator for ubiquitous
  machine-learning,'' \emph{ACM SIGARCH Computer Architecture News}, vol.~42,
  no.~1, pp. 269--284, 2014.

\bibitem{chen2016eyeriss}
Y.-H. Chen, T.~Krishna, J.~S. Emer, and V.~Sze, ``Eyeriss: An energy-efficient
  reconfigurable accelerator for deep convolutional neural networks,''
  \emph{IEEE Journal of Solid-State Circuits}, vol.~52, no.~1, pp. 127--138,
  2016.

\bibitem{chen2019eyeriss}
Y.-H. Chen, T.-J. Yang, J.~Emer, and V.~Sze, ``Eyeriss v2: A flexible
  accelerator for emerging deep neural networks on mobile devices,'' \emph{IEEE
  Journal on Emerging and Selected Topics in Circuits and Systems}, vol.~9,
  no.~2, pp. 292--308, 2019.

\bibitem{10558026}
J.~Choi, W.~Jo, S.~Hong, B.~Kwon, W.~Park, and H.-J. Yoo, ``A 28.6 mj/iter
  stable diffusion processor for text-to-image generation with patch
  similarity-based sparsity augmentation and text-based mixed-precision,'' in
  \emph{2024 IEEE International Symposium on Circuits and Systems (ISCAS)},
  2024, pp. 1--5.

\bibitem{dao2023flashattention2fasterattentionbetter}
\BIBentryALTinterwordspacing
T.~Dao, ``Flashattention-2: Faster attention with better parallelism and work
  partitioning,'' 2023. [Online]. Available:
  \url{https://arxiv.org/abs/2307.08691}
\BIBentrySTDinterwordspacing

\bibitem{dao2022flashattentionfastmemoryefficientexact}
\BIBentryALTinterwordspacing
T.~Dao, D.~Y. Fu, S.~Ermon, A.~Rudra, and C.~Ré, ``Flashattention: Fast and
  memory-efficient exact attention with io-awareness,'' 2022. [Online].
  Available: \url{https://arxiv.org/abs/2205.14135}
\BIBentrySTDinterwordspacing

\bibitem{devlin2019bertpretrainingdeepbidirectional}
\BIBentryALTinterwordspacing
J.~Devlin, M.-W. Chang, K.~Lee, and K.~Toutanova, ``Bert: Pre-training of deep
  bidirectional transformers for language understanding,'' 2019. [Online].
  Available: \url{https://arxiv.org/abs/1810.04805}
\BIBentrySTDinterwordspacing

\bibitem{butter}
H.~Fan, T.~Chau, S.~I. Venieris, R.~Lee, A.~Kouris, W.~Luk, N.~D. Lane, and
  M.~S. Abdelfattah, ``Adaptable butterfly accelerator for attention-based nns
  via hardware and algorithm co-design,'' in \emph{2022 55th IEEE/ACM
  International Symposium on Microarchitecture}.\hskip 1em plus 0.5em minus
  0.4em\relax IEEE, 2022, pp. 599--615.

\bibitem{fang2024structural}
G.~Fang, X.~Ma, and X.~Wang, ``Structural pruning for diffusion models,''
  \emph{Advances in Neural Information Processing Systems}, vol.~36, 2024.

\bibitem{genc2021gemmini}
H.~Genc, S.~Kim, A.~Amid, A.~Haj-Ali, V.~Iyer, P.~Prakash, J.~Zhao, D.~Grubb,
  H.~Liew, H.~Mao \emph{et~al.}, ``Gemmini: Enabling systematic deep-learning
  architecture evaluation via full-stack integration,'' in \emph{2021 58th
  ACM/IEEE Design Automation Conference}.\hskip 1em plus 0.5em minus
  0.4em\relax IEEE, 2021, pp. 769--774.

\bibitem{a3}
T.~J. Ham, S.~J. Jung, S.~Kim, Y.~H. Oh, Y.~Park, Y.~Song, J.-H. Park, S.~Lee,
  K.~Park, J.~W. Lee \emph{et~al.}, ``A\^{} 3: Accelerating attention
  mechanisms in neural networks with approximation,'' in \emph{2020 IEEE
  International Symposium on High Performance Computer Architecture}.\hskip 1em
  plus 0.5em minus 0.4em\relax IEEE, 2020, pp. 328--341.

\bibitem{hartigan1979k}
J.~A. Hartigan, M.~A. Wong \emph{et~al.}, ``A k-means clustering algorithm,''
  \emph{Applied statistics}, vol.~28, no.~1, pp. 100--108, 1979.

\bibitem{he2024ptqd}
Y.~He, L.~Liu, J.~Liu, W.~Wu, H.~Zhou, and B.~Zhuang, ``Ptqd: Accurate
  post-training quantization for diffusion models,'' \emph{Advances in Neural
  Information Processing Systems}, vol.~36, 2024.

\bibitem{hendrycks2016gaussian}
D.~Hendrycks and K.~Gimpel, ``Gaussian error linear units (gelus),''
  \emph{arXiv preprint arXiv:1606.08415}, 2016.

\bibitem{ho2020denoising}
J.~Ho, A.~Jain, and P.~Abbeel, ``Denoising diffusion probabilistic models,''
  \emph{Advances in Neural Information Processing Systems}, vol.~33, pp.
  6840--6851, 2020.

\bibitem{hong2024flashdecodingfasterlargelanguage}
\BIBentryALTinterwordspacing
K.~Hong, G.~Dai, J.~Xu, Q.~Mao, X.~Li, J.~Liu, K.~Chen, Y.~Dong, and Y.~Wang,
  ``Flashdecoding++: Faster large language model inference on gpus,'' 2024.
  [Online]. Available: \url{https://arxiv.org/abs/2311.01282}
\BIBentrySTDinterwordspacing

\bibitem{jouppi2017datacenter}
N.~P. Jouppi, C.~Young, N.~Patil, D.~Patterson, G.~Agrawal, R.~Bajwa, S.~Bates,
  S.~Bhatia, N.~Boden, A.~Borchers \emph{et~al.}, ``In-datacenter performance
  analysis of a tensor processing unit,'' in \emph{Proceedings of the 44th
  annual international symposium on computer architecture}, 2017, pp. 1--12.

\bibitem{judd2016stripes}
P.~Judd, J.~Albericio, T.~Hetherington, T.~M. Aamodt, and A.~Moshovos,
  ``Stripes: Bit-serial deep neural network computing,'' in \emph{2016 49th
  Annual IEEE/ACM International Symposium on Microarchitecture}.\hskip 1em plus
  0.5em minus 0.4em\relax IEEE, 2016, pp. 1--12.

\bibitem{kao2023flat}
S.-C. Kao, S.~Subramanian, G.~Agrawal, A.~Yazdanbakhsh, and T.~Krishna, ``Flat:
  An optimized dataflow for mitigating attention bottlenecks,'' in
  \emph{Proceedings of the 28th ACM International Conference on Architectural
  Support for Programming Languages and Operating Systems, Volume 2}, 2023, pp.
  295--310.

\bibitem{karras2022elucidating}
T.~Karras, M.~Aittala, T.~Aila, and S.~Laine, ``Elucidating the design space of
  diffusion-based generative models,'' \emph{Advances in Neural Information
  Processing Systems}, vol.~35, pp. 26\,565--26\,577, 2022.

\bibitem{kim2023bk}
B.-K. Kim, H.-K. Song, T.~Castells, and S.~Choi, ``Bk-sdm: Architecturally
  compressed stable diffusion for efficient text-to-image generation,'' in
  \emph{Workshop on Efficient Systems for Foundation Models@ ICML2023}, 2023.

\bibitem{kim2021ibertintegeronlybertquantization}
\BIBentryALTinterwordspacing
S.~Kim, A.~Gholami, Z.~Yao, M.~W. Mahoney, and K.~Keutzer, ``I-bert:
  Integer-only bert quantization,'' 2021. [Online]. Available:
  \url{https://arxiv.org/abs/2101.01321}
\BIBentrySTDinterwordspacing

\bibitem{kim2023stackoptimizationtransformerinference}
\BIBentryALTinterwordspacing
S.~Kim, C.~Hooper, T.~Wattanawong, M.~Kang, R.~Yan, H.~Genc, G.~Dinh, Q.~Huang,
  K.~Keutzer, M.~W. Mahoney, Y.~S. Shao, and A.~Gholami, ``Full stack
  optimization of transformer inference: a survey,'' 2023. [Online]. Available:
  \url{https://arxiv.org/abs/2302.14017}
\BIBentrySTDinterwordspacing

\bibitem{cambriconD2024isca}
W.~Kong \emph{et~al.}, ``{Cambricon-D}: Full-network differential acceleration
  for diffusion models,'' in \emph{2024 ACM/IEEE 48th Annual International
  Symposium on Computer Architecture (ISCA)}.\hskip 1em plus 0.5em minus
  0.4em\relax IEEE, 2024.

\bibitem{kung2019packing}
H.~Kung, B.~McDanel, and S.~Q. Zhang, ``Packing sparse convolutional neural
  networks for efficient systolic array implementations: Column combining under
  joint optimization,'' in \emph{Proceedings of the Twenty-Fourth International
  Conference on Architectural Support for Programming Languages and Operating
  Systems}, 2019, pp. 821--834.

\bibitem{kuon2006measuring}
I.~Kuon and J.~Rose, ``Measuring the gap between fpgas and asics,'' in
  \emph{Proceedings of the 2006 ACM/SIGDA 14th international symposium on Field
  programmable gate arrays}, 2006, pp. 21--30.

\bibitem{kwon2018maeri}
H.~Kwon, A.~Samajdar, and T.~Krishna, ``Maeri: Enabling flexible dataflow
  mapping over dnn accelerators via reconfigurable interconnects,'' \emph{ACM
  SIGPLAN Notices}, vol.~53, no.~2, pp. 461--475, 2018.

\bibitem{li2023q}
X.~Li, Y.~Liu, L.~Lian, H.~Yang, Z.~Dong, D.~Kang, S.~Zhang, and K.~Keutzer,
  ``Q-diffusion: Quantizing diffusion models,'' in \emph{Proceedings of the
  IEEE/CVF International Conference on Computer Vision}, 2023, pp.
  17\,535--17\,545.

\bibitem{li2024snapfusion}
Y.~Li, H.~Wang, Q.~Jin, J.~Hu, P.~Chemerys, Y.~Fu, Y.~Wang, S.~Tulyakov, and
  J.~Ren, ``Snapfusion: Text-to-image diffusion model on mobile devices within
  two seconds,'' \emph{Advances in Neural Information Processing Systems},
  vol.~36, 2024.

\bibitem{liao2019davinci}
H.~Liao, J.~Tu, J.~Xia, and X.~Zhou, ``Davinci: A scalable architecture for
  neural network computing.'' in \emph{Hot Chips Symposium}, 2019, pp. 1--44.

\bibitem{lin2014microsoft}
T.-Y. Lin, M.~Maire, S.~Belongie, J.~Hays, P.~Perona, D.~Ramanan,
  P.~Doll{\'a}r, and C.~L. Zitnick, ``Microsoft coco: Common objects in
  context,'' in \emph{Computer Vision--ECCV 2014: 13th European Conference,
  Zurich, Switzerland, September 6-12, 2014, Proceedings, Part V 13}.\hskip 1em
  plus 0.5em minus 0.4em\relax Springer, 2014, pp. 740--755.

\bibitem{liu2022pseudo}
L.~Liu, Y.~Ren, Z.~Lin, and Z.~Zhao, ``Pseudo numerical methods for diffusion
  models on manifolds,'' \emph{arXiv preprint arXiv:2202.09778}, 2022.

\bibitem{lu2022dpm}
C.~Lu, Y.~Zhou, F.~Bao, J.~Chen, C.~Li, and J.~Zhu, ``Dpm-solver: A fast ode
  solver for diffusion probabilistic model sampling in around 10 steps,''
  \emph{Advances in Neural Information Processing Systems}, vol.~35, pp.
  5775--5787, 2022.

\bibitem{lu2021sanger}
L.~Lu, Y.~Jin, H.~Bi, Z.~Luo, P.~Li, T.~Wang, and Y.~Liang, ``Sanger: A
  co-design framework for enabling sparse attention using reconfigurable
  architecture,'' in \emph{MICRO-54: 54th Annual IEEE/ACM International
  Symposium on Microarchitecture}, 2021, pp. 977--991.

\bibitem{lu2017flexflow}
W.~Lu, G.~Yan, J.~Li, S.~Gong, Y.~Han, and X.~Li, ``Flexflow: A flexible
  dataflow accelerator architecture for convolutional neural networks,'' in
  \emph{2017 IEEE International Symposium on High Performance Computer
  Architecture (HPCA)}.\hskip 1em plus 0.5em minus 0.4em\relax IEEE, 2017, pp.
  553--564.

\bibitem{lyu2022accelerating}
Z.~Lyu, X.~Xu, C.~Yang, D.~Lin, and B.~Dai, ``Accelerating diffusion models via
  early stop of the diffusion process,'' \emph{arXiv preprint
  arXiv:2205.12524}, 2022.

\bibitem{ma2023deepcache}
X.~Ma, G.~Fang, and X.~Wang, ``Deepcache: Accelerating diffusion models for
  free,'' in \emph{IEEE/CVF Conference on Computer Vision and Pattern
  Recognition}, 2024.

\bibitem{meng2023distillation}
C.~Meng, R.~Rombach, R.~Gao, D.~Kingma, S.~Ermon, J.~Ho, and T.~Salimans, ``On
  distillation of guided diffusion models,'' in \emph{IEEE/CVF Conference on
  Computer Vision and Pattern Recognition}, 2023, pp. 14\,297--14\,306.

\bibitem{milakov2018onlinenormalizercalculationsoftmax}
\BIBentryALTinterwordspacing
M.~Milakov and N.~Gimelshein, ``Online normalizer calculation for softmax,''
  2018. [Online]. Available: \url{https://arxiv.org/abs/1805.02867}
\BIBentrySTDinterwordspacing

\bibitem{nagel2021white}
M.~Nagel, M.~Fournarakis, R.~A. Amjad, Y.~Bondarenko, M.~Van~Baalen, and
  T.~Blankevoort, ``A white paper on neural network quantization,'' \emph{arXiv
  preprint arXiv:2106.08295}, 2021.

\bibitem{narechania2021vitality}
A.~Narechania, A.~Karduni, R.~Wesslen, and E.~Wall, ``Vitality: Promoting
  serendipitous discovery of academic literature with transformers \& visual
  analytics,'' \emph{IEEE Transactions on Visualization and Computer Graphics},
  vol.~28, no.~1, pp. 486--496, 2021.

\bibitem{nvdla}
Nvidia., ``Nvidia deep learning accelerator.'' \url{http://nvdla.org/}, 2018.

\bibitem{pan2024t}
Z.~Pan, B.~Zhuang, D.-A. Huang, W.~Nie, Z.~Yu, C.~Xiao, J.~Cai, and
  A.~Anandkumar, ``T-stitch: Accelerating sampling in pre-trained diffusion
  models with trajectory stitching,'' \emph{arXiv preprint arXiv:2402.14167},
  2024.

\bibitem{energy}
J.~T. Pawlowski, ``Hybrid memory cube (hmc),'' in \emph{2011 IEEE Hot chips 23
  symposium}.\hskip 1em plus 0.5em minus 0.4em\relax IEEE, 2011, pp. 1--24.

\bibitem{qin2020sigma}
E.~Qin, A.~Samajdar, H.~Kwon, V.~Nadella, S.~Srinivasan, D.~Das, B.~Kaul, and
  T.~Krishna, ``Sigma: A sparse and irregular gemm accelerator with flexible
  interconnects for dnn training,'' in \emph{2020 IEEE International Symposium
  on High Performance Computer Architecture (HPCA)}.\hskip 1em plus 0.5em minus
  0.4em\relax IEEE, 2020, pp. 58--70.

\bibitem{dota}
Z.~Qu, L.~Liu, F.~Tu, Z.~Chen, Y.~Ding, and Y.~Xie, ``Dota: detect and omit
  weak attentions for scalable transformer acceleration,'' in \emph{Proceedings
  of the 27th ACM International Conference on Architectural Support for
  Programming Languages and Operating Systems}, 2022, pp. 14--26.

\bibitem{ramesh2021zero}
A.~Ramesh, M.~Pavlov, G.~Goh, S.~Gray, C.~Voss, A.~Radford, M.~Chen, and
  I.~Sutskever, ``Zero-shot text-to-image generation,'' in \emph{International
  Conference on Machine Learning}.\hskip 1em plus 0.5em minus 0.4em\relax PMLR,
  2021, pp. 8821--8831.

\bibitem{rombach2022high}
R.~Rombach, A.~Blattmann, D.~Lorenz, P.~Esser, and B.~Ommer, ``High-resolution
  image synthesis with latent diffusion models,'' in \emph{IEEE/CVF Conference
  on Computer Vision and Pattern Recognition}, 2022, pp. 10\,684--10\,695.

\bibitem{sauer2024fast}
A.~Sauer, F.~Boesel, T.~Dockhorn, A.~Blattmann, P.~Esser, and R.~Rombach,
  ``Fast high-resolution image synthesis with latent adversarial diffusion
  distillation,'' \emph{arXiv preprint arXiv:2403.12015}, 2024.

\bibitem{sauer2023adversarial}
A.~Sauer, D.~Lorenz, A.~Blattmann, and R.~Rombach, ``Adversarial diffusion
  distillation,'' \emph{arXiv preprint arXiv:2311.17042}, 2023.

\bibitem{sohl2015deep}
J.~Sohl-Dickstein, E.~Weiss, N.~Maheswaranathan, and S.~Ganguli, ``Deep
  unsupervised learning using nonequilibrium thermodynamics,'' in
  \emph{International Conference on Machine Learning}, 2015, pp. 2256--2265.

\bibitem{soltaniyeh2022accelerator}
M.~Soltaniyeh, R.~P. Martin, and S.~Nagarakatte, ``An accelerator for sparse
  convolutional neural networks leveraging systolic general matrix-matrix
  multiplication,'' \emph{ACM Transactions on Architecture and Code
  Optimization (TACO)}, vol.~19, no.~3, pp. 1--26, 2022.

\bibitem{10.1109/ISCA45697.2020.00086}
\BIBentryALTinterwordspacing
Z.~Song, B.~Fu, F.~Wu, Z.~Jiang, L.~Jiang, N.~Jing, and X.~Liang, ``Drq:
  dynamic region-based quantization for deep neural network acceleration,'' in
  \emph{Proceedings of the ACM/IEEE 47th Annual International Symposium on
  Computer Architecture}, ser. ISCA '20.\hskip 1em plus 0.5em minus 0.4em\relax
  IEEE Press, 2020, p. 1010–1021. [Online]. Available:
  \url{https://doi.org/10.1109/ISCA45697.2020.00086}
\BIBentrySTDinterwordspacing

\bibitem{9586134}
J.~R. Stevens, R.~Venkatesan, S.~Dai, B.~Khailany, and A.~Raghunathan,
  ``Softermax: Hardware/software co-design of an efficient softmax for
  transformers,'' in \emph{2021 58th ACM/IEEE Design Automation Conference
  (DAC)}, 2021, pp. 469--474.

\bibitem{sze2020efficient}
V.~Sze, Y.-H. Chen, T.-J. Yang, and J.~S. Emer, \emph{Efficient processing of
  deep neural networks}.\hskip 1em plus 0.5em minus 0.4em\relax Springer, 2020.

\bibitem{tong2024feather}
J.~Tong, A.~Itagi, P.~Chatarasi, and T.~Krishna, ``Feather: A reconfigurable
  accelerator with data reordering support for low-cost on-chip dataflow
  switching,'' \emph{arXiv preprint arXiv:2405.13170}, 2024.

\bibitem{spatten}
H.~Wang, Z.~Zhang, and S.~Han, ``Spatten: Efficient sparse attention
  architecture with cascade token and head pruning,'' in \emph{2021 IEEE
  International Symposium on High-Performance Computer Architecture}.\hskip 1em
  plus 0.5em minus 0.4em\relax IEEE, 2021, pp. 97--110.

\bibitem{wei2017automated}
X.~Wei, C.~H. Yu, P.~Zhang, Y.~Chen, Y.~Wang, H.~Hu, Y.~Liang, and J.~Cong,
  ``Automated systolic array architecture synthesis for high throughput cnn
  inference on fpgas,'' in \emph{Proceedings of the 54th Annual Design
  Automation Conference 2017}, 2017, pp. 1--6.

\bibitem{wimbauer2024cache}
F.~Wimbauer, B.~Wu, E.~Schoenfeld, X.~Dai, J.~Hou, Z.~He, A.~Sanakoyeu,
  P.~Zhang, S.~Tsai, J.~Kohler \emph{et~al.}, ``Cache me if you can:
  Accelerating diffusion models through block caching,'' in \emph{Proceedings
  of the IEEE/CVF Conference on Computer Vision and Pattern Recognition}, 2024,
  pp. 6211--6220.

\bibitem{yang2023denoising}
S.~Yang, Y.~Chen, L.~Wang, S.~Liu, and Y.~Chen, ``Denoising diffusion
  step-aware models,'' \emph{arXiv preprint arXiv:2310.03337}, 2023.

\bibitem{you2023vitcod}
H.~You, Z.~Sun, H.~Shi, Z.~Yu, Y.~Zhao, Y.~Zhang, C.~Li, B.~Li, and Y.~Lin,
  ``Vitcod: Vision transformer acceleration via dedicated algorithm and
  accelerator co-design,'' in \emph{2023 IEEE International Symposium on
  High-Performance Computer Architecture}.\hskip 1em plus 0.5em minus
  0.4em\relax IEEE, 2023, pp. 273--286.

\bibitem{yu2022scaling}
J.~Yu, Y.~Xu, J.~Y. Koh, T.~Luong, G.~Baid, Z.~Wang, V.~Vasudevan, A.~Ku,
  Y.~Yang, B.~K. Ayan \emph{et~al.}, ``Scaling autoregressive models for
  content-rich text-to-image generation,'' \emph{arXiv preprint
  arXiv:2206.10789}, vol.~2, no.~3, p.~5, 2022.

\bibitem{yu2021nnlutneuralapproximationnonlinear}
\BIBentryALTinterwordspacing
J.~Yu, J.~Park, S.~Park, M.~Kim, S.~Lee, D.~H. Lee, and J.~Choi, ``Nn-lut:
  Neural approximation of non-linear operations for efficient transformer
  inference,'' 2021. [Online]. Available:
  \url{https://arxiv.org/abs/2112.02191}
\BIBentrySTDinterwordspacing

\bibitem{zhang2015optimizing}
C.~Zhang, P.~Li, G.~Sun, Y.~Guan, B.~Xiao, and J.~Cong, ``Optimizing fpga-based
  accelerator design for deep convolutional neural networks,'' in
  \emph{Proceedings of the 2015 ACM/SIGDA international symposium on
  field-programmable gate arrays}, 2015, pp. 161--170.

\bibitem{thop}
L.~Zhu, \url{https://github.com/Lyken17/pytorch-OpCounter}, 2018.

\end{thebibliography}
\end{document}